\documentclass[journal,10pt]{IEEEtran}

\usepackage{cite}
\usepackage{hyperref}
\usepackage{amsthm}
\ifCLASSINFOpdf
\usepackage[pdftex]{graphicx}
\else
\fi

\usepackage{amsmath}
\usepackage{tabularx}
\usepackage{booktabs}

\ifCLASSOPTIONcompsoc
  \usepackage[caption=false,font=normalsize,labelfont=sf,textfont=sf]{subfig}
\else
  \usepackage[caption=false,font=footnotesize]{subfig}
\fi

\usepackage{stfloats}
\usepackage{multicol, multirow}
\usepackage{textcomp}
\usepackage{xcolor}
\usepackage{color,soul}
\usepackage{pgfplots}
\usepackage{pgfplotstable}
\usepackage{filecontents}
\usepackage{colortbl}

\usepgfplotslibrary{statistics}

\usepackage{pgfplots}
\usepgfplotslibrary{statistics}
\pgfplotsset{compat=1.8}

\begin{document}

\bstctlcite{IEEEexample:BSTcontrol}

\title{Comparing State-of-the-Art and Emerging Augmented Reality Interfaces for Autonomous Vehicle-to-Pedestrian Communication}


\author{  F.~Gabriele~Prattic\`o,~\IEEEmembership{Student Member,~IEEE,}
        Fabrizio~Lamberti,~\IEEEmembership{Senior Member,~IEEE,}
        Alberto~Cannav\`o,\\~\IEEEmembership{Student Member,~IEEE,}
        Lia~Morra,~\IEEEmembership{Senior Member,~IEEE,}
        Paolo~Montuschi,~\IEEEmembership{Fellow,~IEEE}  
\thanks{Manuscript received XXXX XX, XXXX; revised XXXX XX, XXXX.} 
\thanks{Copyright (c) 2015 IEEE. Personal use of this material is permitted. However, permission to use this material for any other purposes must be obtained from the IEEE by sending a request to \href{mailto:pubs-permissions@ieee.org}{pubs-permissions@ieee.org}. }
\thanks{The authors are with the GRAINS -- GRAphics And INtelligent Systems group at the Dipartimento di Automatica e Informatica of Politecnico di Torino, 10129, Torino, Italy. e-mail: (e-mail: filippogabriele.prattico@polito.it;fabrizio.lamberti@polito.it; alberto.cannavo@polito.it; lia.morra@polito.it; paolo.montuschi@polito.it). 

This article has supplementary downloadable material available at \href{https://doi.org/10.1109/TVT.2021.3054312}{https://doi.org/10.1109/TVT.2021.3054312}, provided by the authors.} }

\markboth{IEEE Transactions on Vehicular Technology,~Vol.~XX, No.~XX, XXXX~2020}%
{Prattic\`o \MakeLowercase{\textit{et al.}}: Comparing State-of-the-Art and Emerging Augmented Reality Interfaces for Autonomous Vehicle-to-Pedestrian Communication}

\maketitle

\begin{abstract}
Providing pedestrians and other vulnerable road users with a clear indication about a fully autonomous vehicle status and intentions is crucial to make them coexist. In the last few years, a variety of external interfaces have been proposed, leveraging different paradigms and technologies including vehicle-mounted devices (like LED panels), short-range on-road projections, and road infrastructure interfaces (e.g., special asphalts with embedded displays). These designs were experimented in different settings, using mockups, specially prepared vehicles, or virtual environments, with heterogeneous evaluation metrics. Promising interfaces based on Augmented Reality (AR) have been proposed too, but their usability and effectiveness have not been tested yet. This paper aims to complement such body of literature by presenting a comparison of state-of-the-art interfaces and new designs under common conditions. To this aim, an immersive Virtual Reality-based simulation was developed, recreating a well-known scenario represented by pedestrians crossing in urban environments under non-regulated conditions. A user study was then performed to investigate the various dimensions of vehicle-to-pedestrian interaction leveraging objective and subjective metrics. Even though no interface clearly stood out over all the considered dimensions, one of the AR designs achieved state-of-the-art results in terms of safety and trust, at the cost of higher cognitive effort and lower intuitiveness compared to LED panels showing anthropomorphic features. Together with rankings on the various dimensions, indications about advantages and drawbacks of the various alternatives that emerged from this study could provide important information for next developments in the field.
 
\end{abstract}


\IEEEpeerreviewmaketitle

\section{Introduction}
\label{sec:introduction}

\IEEEPARstart{A}{dvancements} in the field of automation are continuous, and promise to revolutionize most of everyone's activities. Autonomous vehicles, in particular, will play a key role in this ongoing revolution. While in the early 2010's autonomous vehicles were still regarded as visionary by almost all car manufacturers, today this sector has changed into a multi-billion dollars business. Huge investments are being made\cite{mobility}, and the expectations are that fully autonomous vehicles (FAVs) \cite{Sae} will reach the market by the next decade. Given the disruptive potential of FAVs, their acceptance could be hindered by open challenges related not only to technical aspects, but also to societal factors\cite{xun2019automobile}, which deserve significant attention as both people with and without technical expertise will be requested to trust machines \cite{networking}. This need can be addressed from two viewpoints: that of the drivers (in the future, of the passengers) and of in-vehicle interfaces; and that of vulnerable road users (VRUs), like, e.g., pedestrians, and of interfaces external to the vehicle.

In the last years, significant efforts have been devoted to designing interaction paradigms capable of raising occupants’ awareness about vehicles’ status and intentions, with the aim of improving trust in their autonomous decisions \cite{morra}.  
However, tackling the needs and expectations of VRUs is substantially more complicated. 
Driving is a complex social behavior based on continuous interactions between drivers and road users in uncertain and ambiguous situations.
When adding (or replacing traditional vehicles with) FAVs, the lack of human drivers may cause a communication breakdown, potentially dangerous for all road users \cite{Rasouli_two}. For this reason, interfaces with VRUs (mainly pedestrians) recently started to be investigated to increase the safety and acceptability of FAVs.

Several alternatives have been proposed already, focusing on the most common vehicle-to-pedestrian interactions, i.e., road crossings.  Possible interaction paradigms include showing anthropomorphic features \cite{Chang}, or using LED strips/panels to communicate FAV's intentions \cite{Li, Habibovic, Lagstrom}. 
In other designs, on-vehicle visual hints were replaced by on-road projections, possibly leveraging well-known metaphors like crosswalk or stop signs \cite{HowShould, Nguyen}. Some prototype implementations also introduced changes in the road infrastructure, collecting data from connected vehicles to communicate with pedestrians.

Whenever a new vehicle-to-pedestrian interaction paradigm (or interface) was introduced, it was generally compared with some of the previously proposed ones in qualitative and/or quantitative terms, often working with prototype implementations or mockups. Some works resorted to Virtual Reality (VR) for comparing a number of alternatives at once \cite{HowShould}.
Despite the great relevance of these works for next developments in the field, available experimental evaluations suffer from several drawbacks. For instance, when a representative set of interfaces are considered, not all the experimental conditions studied in other works are investigated (and viceversa).
When the above requirements are met, often the analysis does not fully recreate real-world conditions or does not address the same objective and subjective dimensions which were deemed relevant in other studies. Finally, and most importantly, not all the interaction paradigms devised so far have been tested. This is the case, for instance, of Augmented Reality (AR)-based interfaces, which proved to be particularly effective for in-vehicle interaction \cite{morra}. Some AR-based designs for vehicle-to-pedestrian interfaces have been also proposed but, at present, they have not been compared (or even tested) yet with the mentioned alternatives. 

In this work, a VR-based simulation system was first designed, by taking into account the above issues and endowing it with the capabilities required to support a fair comparison of multiple interface designs relying on heterogeneous technologies and offering different functionalities. This system was then exploited to run a user study aimed to compare the most relevant interfaces proposed so far from different categories. Specifically, AR-based interfaces were considered, to shed some light on their possible role in next-generation vehicle-to-pedestrian interaction paradigms. A wide set of metrics derived from relevant literature was used, providing interested readers with a comprehensive picture of advantages and drawbacks of available paradigms. 

\section{Related work}
\label{sec:related_work}

As discussed in the previous section, the implicit communication mechanism provided by vehicle movement alone might not be able to guarantee an efficient interaction between VRUs and FAVs. 
This is why, in the last years, a number of interfaces have been developed to elicit vehicles' status and/or intentions. 

\subsubsection{Vehicle-mounted interfaces} 
the first interfaces proposed relied on visual hints provided by equipment mounted on the vehicle exterior. A first example is represented by the \textit{``Eyes on a car''} design \cite{Chang}, which aims to replace eye contact between pedestrians and drivers of conventional vehicles with digital eyes placed on the front lights. As soon as the vehicle's sensors detect a pedestrian intending to cross, the eyes start staring at him or her to signal that it will stop to allow crossing. Otherwise, the eyes gaze remains fixed on the road. 
Experiments performed in a Virtual Environment (VE) showed that pedestrians were faster in deciding whether to cross or not when the interface was available. However, some users found the eyes artificial, leading to an undesirable Uncanny Valley effect \cite{HowShould}, and not sufficiently reliable or safe. 

Other interfaces were inspired by the well-known \textit{traffic light}  metaphor. For instance, the concept in \cite{Li} is based on a LED light placed in the vehicle's front. Two lighting patterns were proposed: green, flashing yellow, red (GYR), and white, flashing red, red (WRR). For each set, the first color indicates when it is safe to cross, the last color when it is not; to refer to intermediate situations, the middle color is used. Studies performed on this design confirmed that users were able to properly associate colors to messages the vehicle intended to communicate. However, the flashing red of WRR tended to be (mis)interpreted as a danger signal rather than a warning one, especially compared to the GYR flashing yellow. In another study dealing with color codes \cite{Shinya}, the authors suggested to remove the intermediate warning state (collapsing it to danger), as they realized that  pedestrians tended to selfishly cross even when they were putting the vehicle's passengers at risk. Authors also stated that an indication of approaching vehicle could be more effective than a safe/unsafe to cross information (though without any experimental evaluation).

Another type of interface which proved to be quite intuitive for pedestrians consists in \textit{LED strips} mounted in various positions of the vehicle's exterior. An example is given in \cite{Habibovic, Lagstrom}. In this implementation, the strip is mounted over the windshield, and only the central LEDs are lighted while driving. When a pedestrian in detected, LEDs start to light from the center to the edges of the strip. During crossing, all the LEDs are lighted, and when the vehicle resumes motion, a reverted animation from edges to center is activated. Experiments carried out with a Wizard of Oz technique on a specially prepared vehicle showed that, after a short training, users were able to properly use the interface. It is worth noting that, in \cite{Habibovic}, two different experiments were run to gather both direct and indirect measures of the pedestrian perceived safety. In \cite{HowShould}, a comparison between the above interface and a different design with two strips on the vehicle's sides was performed in a VE. Users preferred the latter interface  showing appreciation for its intuitiveness and unambiguity, though continuous feedback provided was judged as not particularly useful. 

Other examples of this interface category are provided in\cite{Florentine, Lee, autonomi}.
Besides the position on the vehicle's exterior, the main differences among the various designs lay in the strip's shape, in the lighting pattern, and in the color(s) used (and their meanings). In some cases, the above features are combined together. For instance, in \cite{Bockle}, a flashing yellow light is used to indicate that the vehicle is not yielding, whereas a blue light moving from top to bottom indicates that it is going to stop; a fading blue light shows that vehicle is waiting and, lastly, flashing yellow is used again to indicate vehicle restart. 

An alternative design, still based on a LED strip but exploiting again anthropomorphic features like the interface in \cite{Chang}, is the \textit{``Smiling car''} \cite{HowShould}. The interface shows a horizontal yellow line in normal driving conditions, which changes to a smile when a pedestrian is detected to inform him or her that the vehicle will yield. Based on studies performed in a scenario involving one vehicle and one pedestrian \cite{HowShould, Shuchisnigdha}, users found this interface as very simple to use and able to provide unambiguous information.

LED panels have been used also to provide pedestrians with written information on vehicle's status (e.g., ``Braking'' \cite{Shuchisnigdha}) or on what to do (e.g., ``Cross now'' \cite{Milecia, Clerq, Explicit2}). Compared with other interfaces, such a direct communication approach did not prove particularly effective, due to possible readability and language issues \cite{HowShould}.
In some configurations, interfaces request the pedestrians to notify their intention to cross with a specific gesture. In this case, LED strips have been used to inform them that gesture was correctly recognized \cite{Gruenefeld}.

\subsubsection{Projection-based interfaces}
the main limitation of vehicle-mounted interfaces is that they communicate only by means of visual signs placed on the vehicle, which may be poorly visible either because of adverse weather or lighting conditions, or simply because the vehicle is still too far. 

A way pursued to cope with these limitations consists in \textit{projecting} visual indications on the road. For instance, the prototype interface in \cite{Burns}  projects a pattern of parallel lines in front of the vehicle. Lines are perpendicular to the driving direction, and get closer/farther when the vehicle is decelerating/accelerating. Experiments performed with a real, non-autonomous vehicle showed that pedestrians were faster in deciding whether to cross or not, but they focused their attention on the road rather than on the vehicle itself. In \cite{Nguyen}, a more sophisticated projection pattern was used. During normal driving, a wave-shaped red pattern is projected. When a pedestrian is detected and vehicle starts to slow down, pattern color changes to yellow. When vehicle has come to a complete stop and crossing is safe, pattern changes into a green crosswalk shape, which turns to red when vehicle is going to restart again. The implementation proposed in \cite{HowShould}, characterized by a lower number of states, was rated as very pleasant; it was also  judged as futuristic, but this result could be due to the particular vehicle the interface was mounted onto. Despite the increased visibility, these interfaces may not work well with all road conditions. Moreover, even though they rely on the well-known crosswalk concept which should be familiar for pedestrians, experiments indicated that these interfaces induce a high mental workload.

\subsubsection{Smart road interfaces}

a different approach to support vehicle-to-pedestrian communication consists in using the road infrastructure itself. In the so called ``Smart roads'' concept, visual hints are provided through LED panels embedded in the road pavement to indicate when it is safe to cross (e.g., by showing a white crosswalk) and when it is not (e.g., lighting red bars on the sidewalk edge). For instance, the prototype implemented in London \cite{umbrellium} was judged as very trustworthy. The main drawback of such interfaces is the cost associated with updating the road infrastructure \cite{HowShould}. 

\subsubsection{Multi-modal interfaces}
the above designs rely only on the visual communication channel, which is also the focus of the present study. It is worth noticing, however, that this choice makes these interfaces not suitable for all VRUs, e.g., for visually impaired persons. For this reason, multi-modal interfaces including a combination of visual, audio and haptic-based notifications have been proposed \cite{Mahadevan}. However, the audio channel proved not particularly effective in noisy traffic. 
\subsubsection{AR interfaces}
the last category in this review is represented by AR interfaces. Thanks to  technological advancements and the ever-larger availability of consumer-grade devices \cite{relay}, it is possible to imagine a not too distant future in which pedestrians will wear their own AR glasses,  enabling the development of sophisticated vehicle-to-pedestrian and, more generally, human-robot interaction paradigms \cite{mixedcann}. However, experiments in this field are still rare. In a recent study \cite{Hesenius}, three concepts were explored. In the first one, AR visual hints are exploited to show to the pedestrian, through a blue tape overlapped to his or her field of view, the safer path to follow for crossing the road; moreover, AR-based yellow arrows are drawn in front of the vehicle to show where it could stop, if needed. The second design, referred to as ``safe zones'', uses AR to draw large green regions indicating where it is safe to cross, and red regions closer to the vehicle indicating where it would be dangerous to cross. The third design is a combination of the previous ones. A user study was performed, in which participants were shown static representations of the three designs. Despite the amount of visual indications provided, participants preferred the third design.

The above review, which showed that various interface categories have indeed been proposed but only a few of them have been tested under representative and/or comparable conditions, motivated the design of the simulation system and the experimental analysis that are reported in this work.

\section{Simulation system}
\label{sec:simulation}

In this section, the VR-based system created to support the study of the various interfaces is illustrated. In particular, the simulation environment is presented first. Afterwards, interfaces implementation and scenario configuration are discussed. 

\subsection{Simulation environment}
\label{sec:simulation_env}

The simulation environment has been built on top of the AirSim open-source software \cite{AirSim}. 
Originally developed by Microsoft as a simulator for collecting data needed to train computer vision algorithms for unmanned vehicles, AirSim supports hardware-in-the loop simulations and is characterized by an extremely high visual fidelity. Also, a ``Windridge City'' urban scenario is provided free of charge. 

However, the ``as-is'' performance of AirSim is not suitable for an effective fruition via immersive VR. Hence, data collection from vehicle's simulated sensors was disabled (as not required in this work), and a number of optimizations were implemented in order to target the minimum framerate required for immersive experiences (90 frames per seconds or more to prevent motion sickness). To this purpose, among the graphics platforms supported by AirSim, Unity was selected; optimizations leveraged ad-hoc functionalities available in this graphics engine to improve the performance of the application logic and of graphics processing. 

In particular, concerning the former aspect, the novel Data-Oriented Technology Stack (DOTS) paradigm was used, which allows developers to exploit the parallel processing capabilities and large cache availability of modern processors. The benefit in performance is especially relevant for applications that need to handle multiple instances of the same simulated object, in this case the logic (and physics) of vehicles (and their parts, like wheels) and of pedestrians. 
The above paradigm was implemented by relying on the Unity's architectural pattern named Entity Component System (ECS) together with the C\# Job System, which enables an effective handling of frame-synchronized multi-threading, and the Burst Compiler, which allows for the generation of high-performance native code that can exploit parallel hardware acceleration.

For the second aspect, the Unity's Scriptable Render Pipeline (SRP) was exploited, since it allows to better customize graphics quality based on hardware capabilities and application requirements. In particular, the High Definition Render Pipeline (HDRP), which has been recently made available for VR applications, was used. The HDRP allowed to effectively manage the rendering of 3D elements with different levels of detail (LODs) based on actual distance from the user, to use dynamic resolution for reducing the workload on the GPU, to enable efficient calculation of reflections, transparencies and shadows, and to activate volumetric lighting for realistic effects. Furthermore, the SRP supports the Shader Graph (SG), which is a visual shader programming tool that was of paramount importance to effectively backing the implementation of the vehicle's interfaces.

Finally, several minor changes were applied to the original urban scenario like, among others, simplifying the colliders' shape, activating GPU instancing on materials, and enabling asynchronous map loading.

\subsection{Interface selection and implementation}
\label{sec:interface_implementation}

\begin{figure}%
\centering

\subfloat[Baseline,  B]{\includegraphics[width=1.57in]{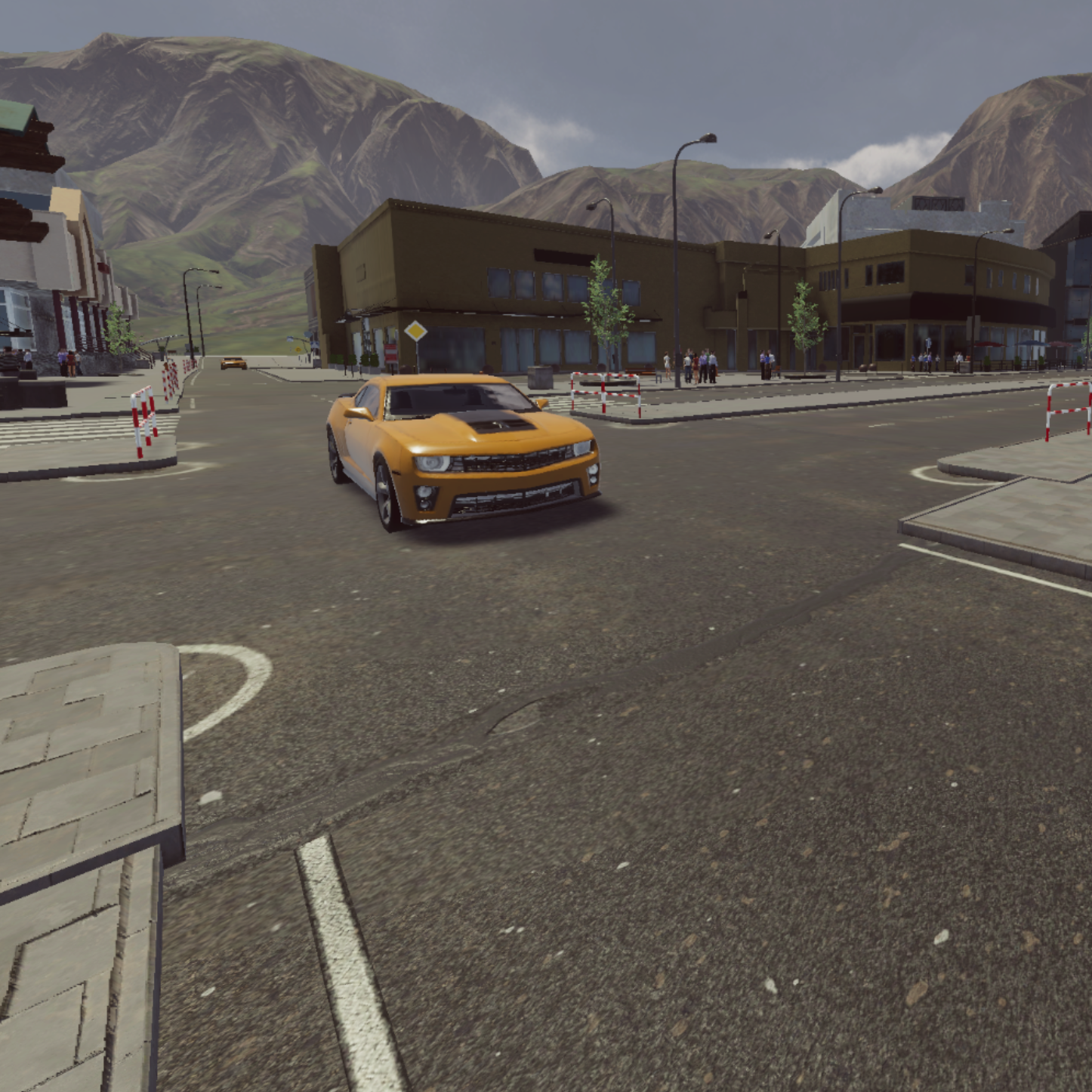}%
\label{fig:baseline}}
\hfil
\subfloat[Smile,  S]{\includegraphics[width=1.57in]{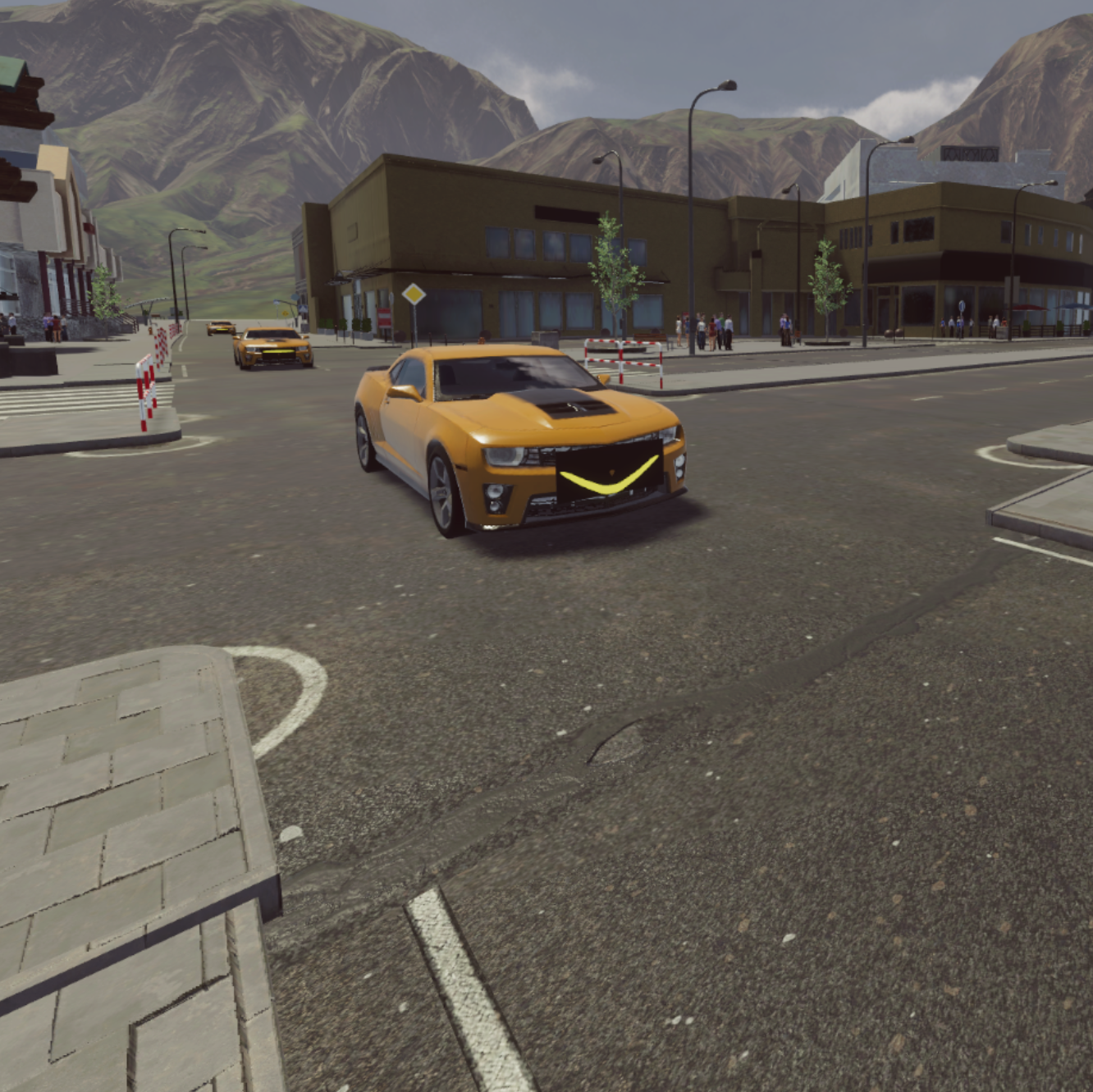}%
\label{fig:smile}}

\subfloat[Projection,  P]{\includegraphics[width=1.57in]{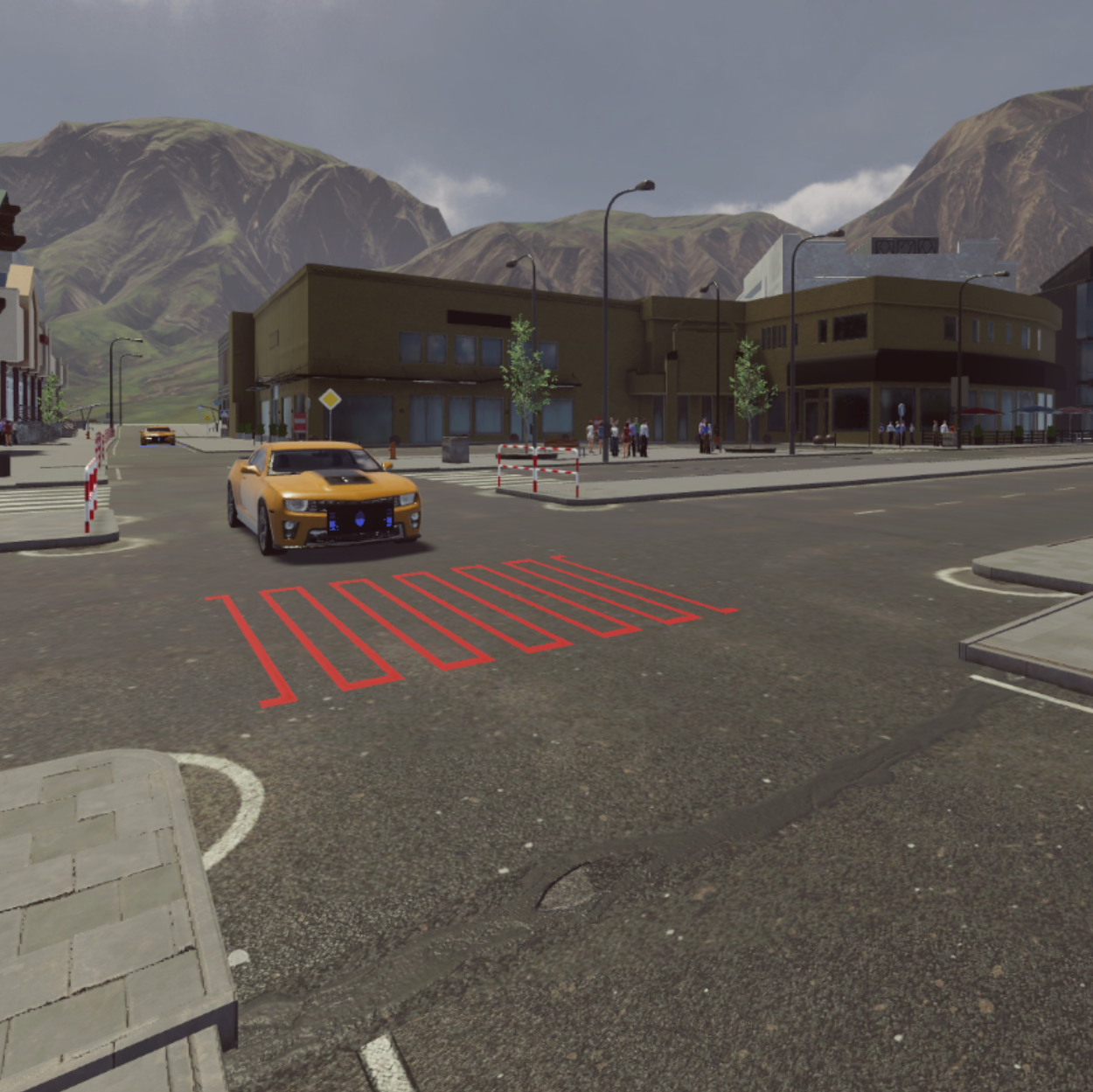}%
\label{fig:projected}}
\hfil
\subfloat[Smart road,  M]{\includegraphics[width=1.57in]{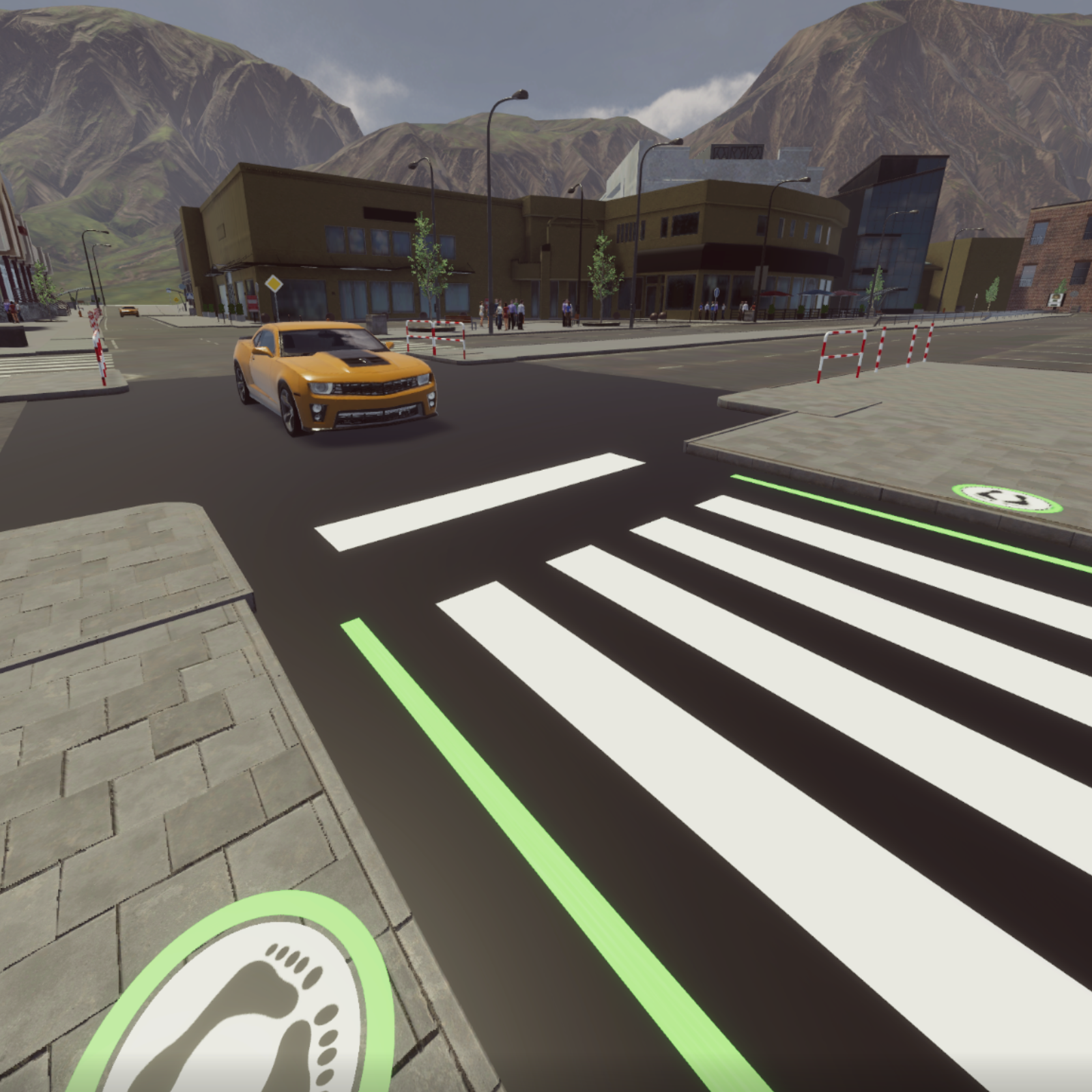}%
\label{fig:smart_road}}

\subfloat[Safe roads,  F]{\includegraphics[width=1.57in]{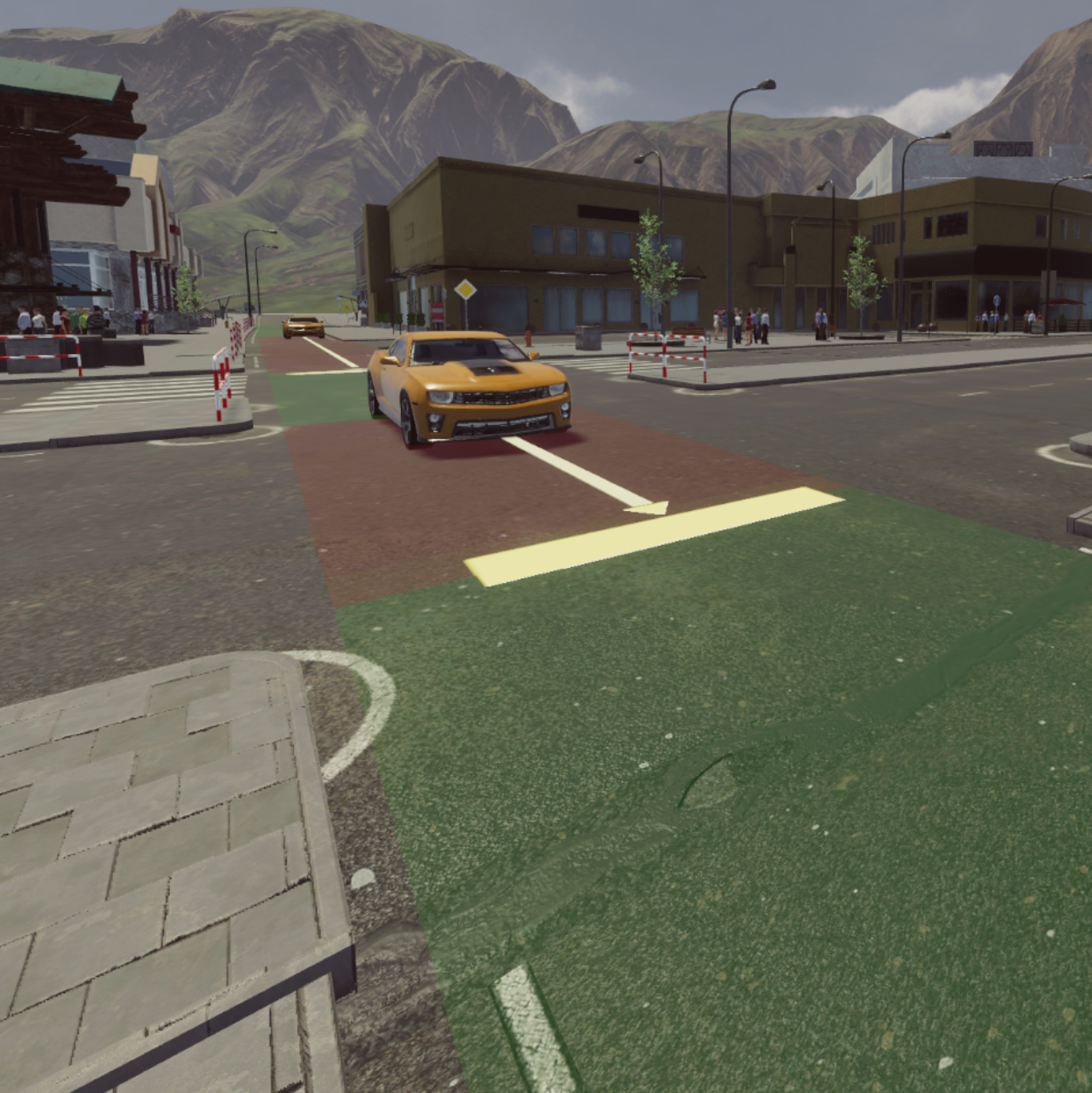}%
\label{fig:safe_road}}
\hfil
\subfloat[Safe roads extended,  E]{\includegraphics[width=1.57in]{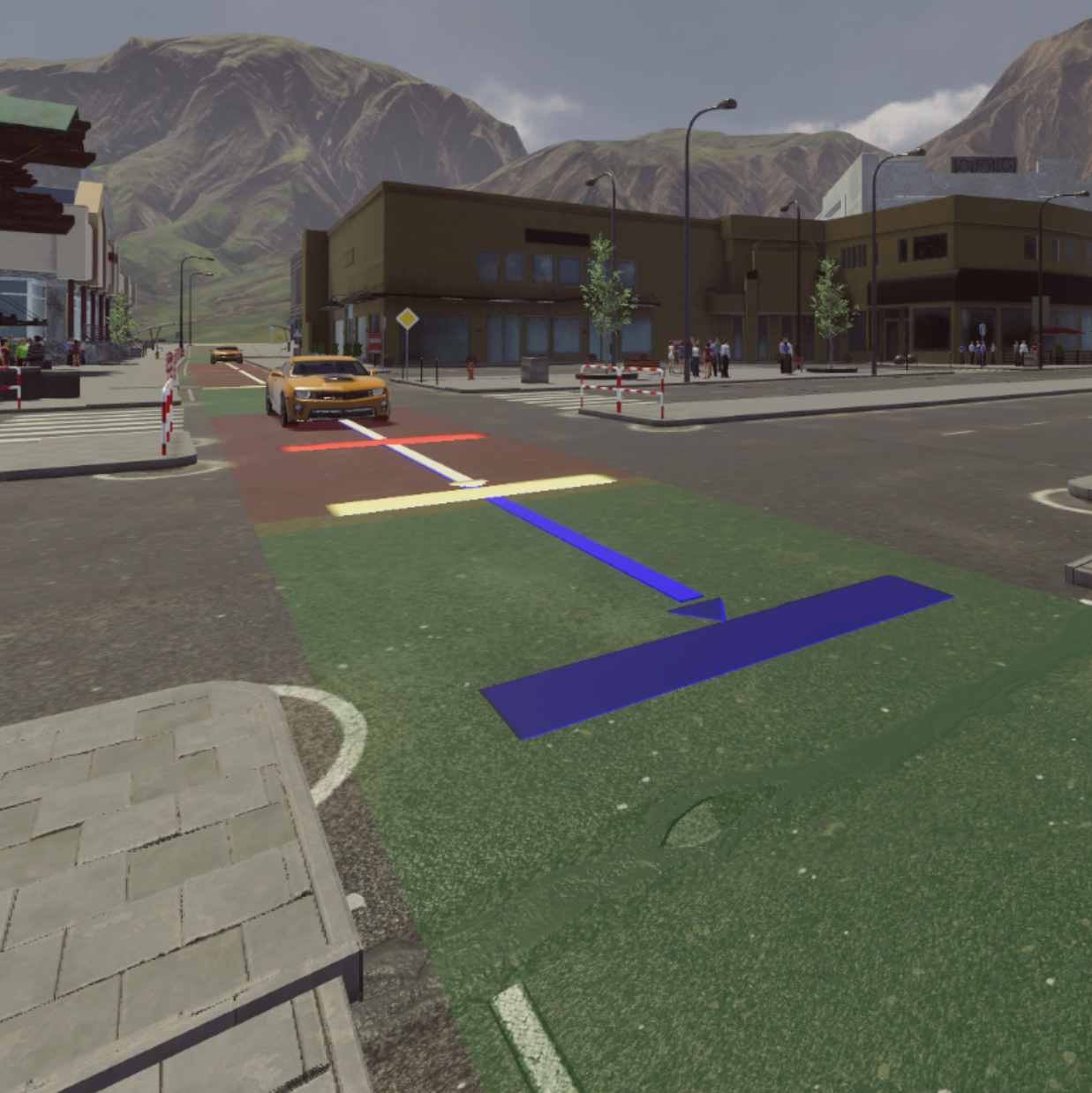}%
\label{fig:safe_road_extended}}

\caption{Simulation environment and interfaces included in the evaluation depicted while the vehicle is braking.}
\label{fig:interfaces}
\end{figure}

In order to provide the users with a reference implementation, a ``Baseline'' behavior was defined for the virtual FAVs, in which no specific interface is available (Fig.~\ref{fig:baseline}). 

Braking was implemented as follows: as soon as the FAV detects a pedestrian intending to cross, because close enough to the sidewalk edge, looking either at the vehicle or at the road, and within the detection range,
 it starts braking with a constant deceleration. Deceleration is calculated by considering the current speed and aiming to reach a full stop at a certain distance from the pedestrian;
in preliminary experiments, other deceleration strategies were found to largely alter the users' perception of the interface behavior, as confirmed also by \cite{frenatissime}, and hence were excluded from the evaluation. If the FAV is not able to stop in due time (too fast and/or too close when crossing starts), the horn is activated to signal the danger. 

By default, AirSim's built-in vehicles are provided with a controller logic for driving them with keyboard or joystick. However, in this work FAVs' behavior described above was scripted using the ECS, and speed was managed using a PID controller (with output shaping) since, as it will be shown later, only deceleration and acceleration on a straight path had to be handled in the designed experiments.

By building on top of the above ``Baseline'' (abbreviated B), one interface was implemented for each of the categories discussed in detail in Section 2, focusing specifically on visual interfaces. To this aim, we chose the interfaces that scored better in experiments reported in previous literature.  Neither the FAV's behavior nor the vehicle shape were altered to avoid biases in the interface comparison \cite{HowShould}.

For vehicle-mounted interfaces \cite{HowShould, Shuchisnigdha}, the ``Smiling car'' concept was selected \cite{HowShould}, later referred to as ``Smile'' (abbreviated S). In normal driving conditions, an horizontal straight yellow line is shown on the vehicle's front side; when a pedestrian is detected and the FAV starts braking, the line turns smoothly into a smile (Fig.~\ref{fig:smile}) to signal that it will yield.

Among projection-based interfaces \cite{HowShould, Nguyen, Burns}, the concept originally presented in \cite{Nguyen} was considered (later named ``Projection'', P). To foster comparability of results, the implementation in \cite{HowShould} was used (Fig.~\ref{fig:projected}),
but the sound played at vehicle restart was removed in order to make the interface rely only on visual indications, like the other ones. The interface integrates a LED panel on the vehicle's front side, whose pattern changes based on actual projection: all LEDs lighted in normal driving conditions, LEDs lighting from edges to center while decelerating, LEDs lighting in the crossing direction while stopped, then transition to the original state at vehicle restart. 

For the ``Smart road'' (M) category, the implementation studied in \cite{umbrellium} was selected (Fig.~\ref{fig:smart_road}). In the original work, pedestrian detection is performed by the infrastructure (thus, for instance, crosswalk marking appears even when there is no vehicle approaching); in this study, detection  was moved onto the FAV’s side. When there is no vehicle approaching, the interface does not provide any feedback (like with the other designs). When a vehicle is approaching, the interface's state is controlled by the vehicle itself, simulating a connection established between the two entities; if the vehicle is able to stop, a crosswalk appears together with the predicted stop position, and white lines on the sidewalk edge are turned green (otherwise, they are turned red and no crosswalk is shown).

Lastly, for the AR-based category, the third design in \cite{Hesenius} leveraging AR hints for showing zones suitable/unsuitable for crossing was chosen, later referred to as ``Safe roads'' (F).
The interface shows in front of the vehicle a yellow arrow, terminated by a yellow bar whose length is equal to the current estimated stopping distance (Fig.~\ref{fig:safe_road}).
The arrow length is determined only by the vehicle's dynamics, and it is not related in any way to pedestrian detection. When accelerating or moving at cruise speed, the stopping distance is calculated by applying a reference deceleration (deemed comfortable for the passengers) of 3 m/s\textsuperscript{2} \cite{Gunnar}. When braking, the stopping distance is obtained by sampling a pre-calculated braking curve at the current velocity: the curve is selected among two alternatives (comfortable 3 m/s\textsuperscript{2}, and maximum 6 m/s\textsuperscript{2}), picking the one with the lowest deceleration value still greater than current deceleration.
To reinforce the feedback provided by the interface, the region of the road between the vehicle and the arrowhead is colored red, the rest green. 
It is worth observing that, since the arrowhead (and the colored region) reflects solely the dynamics of the FAV, the pedestrian is not provided with a clear, immediate indication of the fact that the vehicle actually detected him or her, and will stop accordingly. 

Since the goal of the present work was to study, in particular, the effectiveness of AR-based vehicle-to-pedestrian interaction, a further interface was added to the analysis. This choice was motivated by the fact that, unfortunately, the interface in \cite{Hesenius} is one of the few concepts proposed so far that has still to be tested by letting users actually experiment road crossing with it.  
Hence, a new interface was designed,  named ``Safe roads extended'', by combining key communication abilities that were found in some of the best interface designs, but were lacking in F. 
The behavior of the original yellow arrow is maintained. However, a red tick is added on the arrow body (Fig.~\ref{fig:safe_road_extended}) to show the pedestrian where the vehicle would stop in case of an emergency brake (6 m/s\textsuperscript{2}): the distance between the vehicle and the tick represents the minimum stopping distance. 
Moreover, an additional blue arrow, with a blue bar on the head, is drawn only when a pedestrian is detected (thus avoiding unneeded visual clutter). The blue arrowhead position is fixed as the vehicle's speed decreases, showing the (estimated) point where it will stop; thus, the pedestrian is implicitly informed that he or she has been detected, and can use this information to decide whether to cross or not.

\subsection{Virtual scenario}
\label{sec:virtual_scene}

In order to test the interfaces, a representative case study within the selected VE had to be defined. 
Based on previous literature, we decided to focus on an unregulated crossing scenario \cite{Clerq,HowShould}, which could become commonplace in urban contexts populated only by FAVs. Moreover, in the absence of other external signals (e.g., traffic lights), VRUs have to rely on information provided by the FAVs' interfaces only, leveling out possible environmental contributions. In particular, a one-way, 5 m wide road was chosen, as done, e.g., in \cite{Nguyen, HowShould}. In this way, the users could ground their decisions to cross or not on the observation of a limited number of vehicles moving on a straight path rather than, e.g., on vehicles changing lanes or stopping at different locations; evaluation metrics can also be made independent of the crossing direction. 

\begin{figure*}%
    \centering
    \includegraphics[width=1.7\columnwidth]{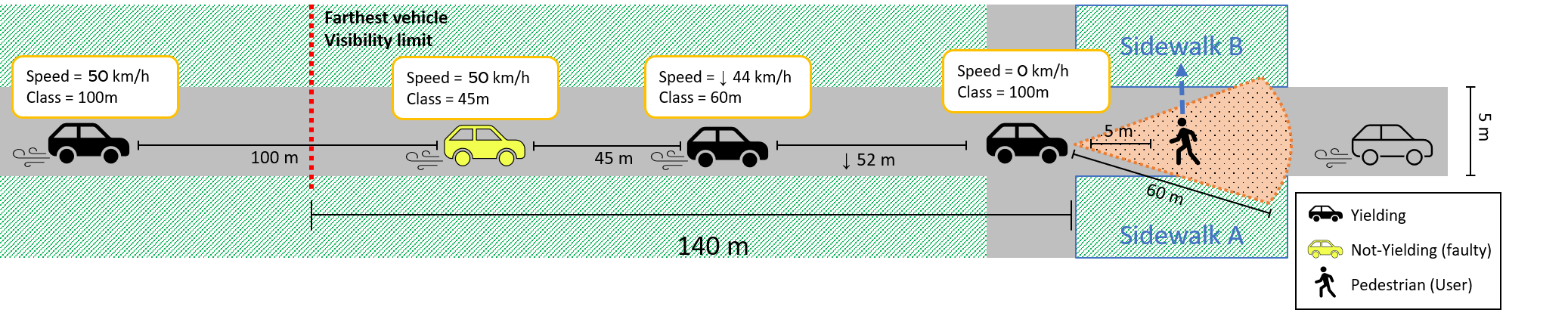}
    \caption{Scenario devised for the experiment: pattern used for generating vehicles, visibility and detection ranges, safety stop distance, and crosswalk length.}
    \label{fig:scene}
\end{figure*}

Following the approach adopted, e.g., in \cite{Clerq, Doric}, vehicles were organized in a pattern whose characteristics were controlled in order to ensure realism while at the same time presenting the users a number of different situations stimulating the various interfaces in many ways (Fig.~\ref{fig:scene}).  
Cruise speed was set to  50 Km/h (14 m/s) for all the vehicles, since previous work indicates that above 40 Km/h the demographics have low to no impact on the pedestrians' behavior \cite{Beggiato}. Since the experience was expected to strongly depend on space (hence, time) separating the users from the approaching vehicle, it was decided to vary the distances between consecutive vehicles, setting them to either 45 m, 60 m or 100 m; within the pattern, a pseudo-random distribution guaranteeing a uniform distribution of the inter-vehicle distance was used. So called ``not-yielding'' vehicles were also included in the pattern: this characteristic has been used rarely in the literature, but authors of works where it was not exploited lamented the fact that its lack negatively influenced the realism of the simulation \cite{HowShould}.
These vehicles are essential to explore trust in human-to-vehicle interactions in the presence of faulty FAVs. Detection range of vehicles' sensors was set to 60 m, and yielding vehicles were programmed to stop 5 m ahead of the crossing pedestrian. 

\section{Experimental protocol}
\label{sec:experimental}

In this section, the experimental protocol devised to carry out the evaluation will be illustrated in detail.

\subsection{Experiment design and preparation}
\label{sec:pre_experiment}

For each interface, the user is instructed to cross as quickly and as many times as possible, whenever he or she feels it safe to cross. The testing of each experience is concluded when the following two conditions are met: the user has completed at least 15 crossings in total and at least one per distance. This choice ensures that the user experiences the full spectrum of situations within a reasonable amount of time. Additionally, as done in \cite{Clerq} a maximum limit to the duration of the simulation was introduced, in our case an upper bound of 300 generated vehicles was set.
FAVs are set to become visible at 140 m (Fig.~\ref{fig:scene}) and are allowed to queue, though queues are prevented from growing too much in order to avoid  long waiting times that could be physically and/or mentally demanding in VR. The user was informed that crossings in front of queued vehicles would not be considered (they will be later referred to as invalid crossings).

As done in almost all the works carried out so far, the user is controlling the only pedestrian that can interact with vehicles in the above scenario: in this way, it is possible to isolate the contribution of his or her interactions with the approaching vehicles from other possibly disturbing factors. It is worth observing that interactions in more complex scenarios could be investigated in the future using the devised software, which already integrates a traffic simulation system encompassing both vehicles and pedestrians (not used in the experiments).

Experiments were designed according to a within-subjects logic: all the subjects tested all the interfaces, starting with the Baseline. Latin Square order was then used to define the sequence of the remaining tests.  
In the future, it is reasonable to expect that pedestrians will be accustomed with FAVs' interfaces: we sought to replicate a similar condition in our experimental setting by showing each subject a video to familiarize with each interface prior to the virtual experience. In each video, a pedestrian attempts to cross the road under three different conditions: the vehicle is either too close to stop (hence, it will pass by horning at the pedestrian), it is forced to an emergency brake, or it has sufficient time to stop smoothly. 
Videos shown to participants are made available\footnote{Videos: \href{https://youtu.be/RoPURY1dlZE}{https://youtu.be/RoPURY1dlZE}}.

\subsection{Hardware configuration and physical setup}
\label{sec:hardware}

The VR headset device selected to let the users immerse themselves in the created VE was the Samsung Odyssey. It is an HMD equipped with an AMOLED display with a resolution of 1440$\times$1600 pixels per eye, a refresh rate of 90Hz and a horizontal field of view of 110$^{\circ}$. The headset supports 3D audio and relies on inside-out tracking technology, meaning that it does not requires external equipment to determine the user's (better, his or her head's) and the controllers' pose in the real environment. 
An untethered setup was defined using the MSI VR One Backpack PC to run the simulation. The backpack integrates an Intel Core i7-7820HK and a NVidia GeForce GTX 1070 in less than 3.5 Kg. Users could thus walk in a physical space mapping one-to-one with the selected portion of the VE where they were expected to cross. This setup was expected to lead to a more realistic experience,  boosting the sense of immersion and presence \cite{letvr}.

\subsection{Objective evaluation metrics}
\label{sec:obj_metrics}

During the simulation, user's behavior is logged by collecting several quantitative measures. 
For each interaction with an approaching vehicle, the system records the time at which the pedestrian is detected (and the vehicle starts braking, thus initiating the negotiation) and the time at which the pedestrian reaches the opposite sidewalk, i.e., the crossing ends (Fig.~\ref{fig:scene}).
\emph{Crossing time} (\textit{CT}), calculated as the difference between the above times, is logged;  previous works  speculated that it may be associated with user's uncertainty \cite{HowShould}. %
When the user enters the road, the distance of the approaching vehicle (\emph{distance at crossing}, or \textit{DAC}) and its speed (\emph{speed at crossing}, or \textit{SAC}) are recorded. The interval between the time the user was detected and he or she left the sidewalk (and negotiation ends) is defined as \emph{decision time} (\textit{DT}). The system also keeps track of possible collisions with the vehicles. Aborted crossings, occurring when the user gets off and on the same sidewalk (without reaching the opposite one), are also recorded. An \textit{efficiency} metric was defined as the total number of valid crossings (i.e., non-invalid and non-aborted) divided by the time elapsed from the first to the last crossing.

\subsection{Subjective evaluation metrics}
\label{sec:subj_metrics}

To further evaluate users' experience, a questionnaire\footnote{Questionnaire: \href{http://tiny.cc/6xoksz}{http://tiny.cc/6xoksz}} was developed based on previous literature on the subject.  

A before-experience section (\textit{BEQ}) collects demographic information about the participant, as well as his or her opinion about FAVs, crossing habits, 
and experience with technologies used in the experiments. 
Possible symptoms associated with motion sickness are recorded using the Simulation Sickness Questionnaire (SSQ) \cite{ssq}. 

Afterwards, the testing of the individual interfaces begins. 
The after-video tutorial section (\textit{ATQ}) aims at ensuring that the interface functioning has been fully understood prior to the actual VR experience: the user is required to answer some questions, describe the interface behavior in words and sketch a graphical representation on paper. Then, the after-interface questionnaire (\textit{AIQ}) is used to investigate various dimensions of vehicle-to-user interaction after the VR experience (lasting approximately 10 minutes). 
The AIQ combines questions from the Trust Scale (TS) \cite{trust_scale}, the  System Usability Scale (SUS) \cite{SUS}, the NASA Task Load Index (NASA-TLX) 
\cite{tlx} and the Short User Experience Questionnaire (UEQ-S) \cite{ueqs}. 
Additional, custom questions are used to, e.g., investigate the level of perceived safety associated with a given interface, including whether the participant has felt the need to wait for the vehicle to come to a full stop before attempting to cross.  Finally, the participant is requested to rate a number of features of the tested interface (perceived safety, familiarity, workload, etc.) with respect to the Baseline. 

After testing all the interfaces, a post-experience questionnaire (PEQ) is administered, which includes questions from the SSQ to evaluate possible discomforts due to the VE, from the VRUse tool \cite{vruse} to determine the usability of the VR simulation, and from the iGroup Presence Questionnaire (IPQ) \cite{ipq} to measure immersion and presence. Additional questions verify that the proper level of realism is reached in the simulation.  Finally, the participant is asked to rank (without ties) all the interfaces based on individual features and overall. A final open interview concludes the experience.

\subsection{Data analysis}
\label{sec:data_analysis}
Collected data were analyzed using MS Excel with the Real-Statistics add-on (v7.1). Comparative analyses were performed using Friedman's test (pass condition was set at $p \leq 0.05$). Post-hoc comparison on pairwise groups was further applied using the Conover's test. For non-comparative questionnaire items, Cronbach's Alpha ($\alpha$) on clusters of related items was computed to test internal consistency. Interface rankings in the PEQ were aggregated using a Bucklin voting system; in particular, the relative placement scoring system (RPSS) \cite{rpss} was applied considering a majority of half of the sample size.

\section{Results}
\label{sec:results}

Experimental evaluation involved 12 volunteers (8 males, 4 females) aged between 24 and 46 ($M=28.00,\,SD=4.68$). Participants were selected among Politecnico di Torino students and from the authors' social networks. Based on demographic information collected from the BEQ, $17\%$ of the participants were unfamiliar with immersive VR (never used, or used once), whereas $83\%$ said they use this technology very often or on a daily basis. 
Participants reported crossing the roads under non-regulated conditions either very often/daily (50\%) or occasionally (50\%), and were on average trustworthy about FAVs ($M=3.42,\,SD=0.78,\,\alpha=0.77$).
In the following, experimental results will be presented, focusing first on the participants' perception of the VE and of simulation quality, since possible discomforts associated to the use of VR could introduce biases in the evaluation. Afterwards, their experience with the various interfaces will be compared by leveraging subjective feedback and objective measurements.

\subsection{Virtual experience and simulation quality}
\label{sec:ve_sim}

No significant effect was registered on pre-post experience conditions ($p$-value computed using the two-tailed Mann-Whitney U-test) for the three SSQ clusters, i.e., oculomotor ($O$), disorientation ($D$), and nausea ($N$) symptoms, as well as overall ($T$): $\Delta\overline{O}=13.3\pm18.6\:(p=0.08)$, $\Delta\overline{D}=8.1\pm15.5\:(p=0.48)$, $\Delta\overline{N}=14.3\pm37.4\:(p=0.51)$, and $\Delta\overline{T}=14.3\pm22.4\:(p=0.07)$. Hence, psychophysiological alterations related to motion sickness that could have affected participants' attention level, reaction time, etc. apparently had no significant influence on the experiments.

Results from the IPQ shows that participants experienced a fairly good general sense of presence in the VE ($M=4.0,\,SD=0.58$), and assigned medium-high scores for the remaining indicators, i.e., spatial presence ($M=3.56, SD=0.46, \alpha=0.80$), involvement ($M=3.10,\,SD=0.66,\,\alpha=0.76$), and realism ($M=3.10,\,SD=0.50,\,\alpha=0.78$). 

Participants were also satisfied with the operation of the locomotion technique  ($M=4.42,\,SD=0.50,\,\alpha=0.79$), as well as with the visual quality and the simulation fidelity of the VE ($M=4.08,\,SD=0.50,\,\alpha=0.72$). 
\begin{table}[]
\centering
\caption{FAV simulation perception. Inverted items are marked with *.}
\label{tab:fav_perc}
\resizebox{0.49\textwidth}{!}{%
\begin{tabular}{lll}
    \toprule
    Item & M & SD \\ \midrule
    {\begin{tabular}[c]{@{}l@{}}You think that vehicles were driven  by a human (1) \\ or were fully autonomous (5)\end{tabular}} & 3.75 & 0.72 \\ 
    The FAVs showed adequate decision making skills & 4.17 & 0.56 \\ 
    {\begin{tabular}[c]{@{}l@{}}The FAVs struggled in case of sudden, unexpected\\or abrupt pedestrian behavior*
        \end{tabular}} & 2.08 & 1.04 \\
    The FAVs made mistakes* & 2.08 & 0.96 \\ 
    The FAVs seemed intelligent & 3.83 & 0.56 \\ 
    Overall, you are satisfied about FAV simulation & 4.00 & 0.59 \\ 
    \bottomrule
\end{tabular}%
}
\end{table}
Scores related to the perception of the FAVs, as measured by the PEQ, are reported in Table~\ref{tab:fav_perc} ($\alpha=0.76$). 
Participants were satisfied overall with the FAV simulation. Based on open feedback collected after the experiments, lower scores assigned to some  questions were due to the non-yielding behavior of some vehicles which, as said, was intentionally introduced to simulate failures in pedestrian detection: in this respect, $75.0\%$ of the participants stated that they felt awkward when, right in the middle of the crossing, the vehicle sometimes did not recognize them (and they ascribed this behavior to faulty, or not  smart enough, FAVs). Few participants reported also that they found FAVs too polite with pedestrians compared to how a human driver would behave in such situations. 

\subsection{Interface comparison}
\label{sec:interface_comp}

Once the representativeness of the simulated scenario was validated, the analysis moved to comparing selected interfaces. 

\subsubsection{Subjective results}

\begin{table*}[]
\centering
\caption{Rankings and p-values for pairwise comparisons. Aggregated ranking is calculated by combining individual feature rankings (overall excluded). Remaining rankings are obtained directly from PEQ answers. Inverted items are marked with *.}
\label{tab:post_rank}
\resizebox{\textwidth}{!}{%
\begin{tabular}{lllllll|lllllllllllllll}
\toprule
                                & B & S & P & M & F & E & B-S & B-P & B-M & B-F & B-E & S-P & S-M & S-F & S-E & P-M & P-F & P-E & M-F & M-E & F-E \\ \midrule
Overall                         & 6 & \cellcolor[HTML]{c0c0c0}2 & 3 & 4 & 5 & \cellcolor[HTML]{656565}1 & \textbf{0.00} & \textbf{0.00} & \textbf{0.04} & \textbf{0.02} & \textbf{0.00} & 0.68 & \textbf{0.02} & \textbf{0.03} & 0.50 & \textbf{0.05} & \textbf{0.07} & 0.28 & 0.84 & \textbf{0.00} & \textbf{0.01} \\
Safety                          & 6 & \cellcolor[HTML]{c0c0c0}2 & 3 & 4 & 5 & \cellcolor[HTML]{656565}1 & \textbf{0.00} &\textbf{0.00} &\textbf{0.00} & \textbf{0.01} &\textbf{0.00} & 0.58 & 0.21 & \textbf{0.04} & 0.40 & 0.07 & \textbf{0.01} & 0.78 & 0.40 & \textbf{0.04} & \textbf{0.00} \\
Familiarity                     & \cellcolor[HTML]{656565}1 & 3 & 4 & \cellcolor[HTML]{c0c0c0}2 & 5 & 6 & \textbf{0.00} &\textbf{0.00} & \textbf{0.04} &\textbf{0.00} &\textbf{0.00} & 0.22 &\textbf{0.00} & \textbf{0.02} &\textbf{0.00} &\textbf{0.00} & 0.22 & \textbf{0.00} & \textbf{0.00} & \textbf{0.00} & \textbf{0.01} \\
Intuitiveness                     & 6 & \cellcolor[HTML]{656565}1 & 3 & \cellcolor[HTML]{c0c0c0}2 & 5 & 4 & \textbf{0.00} & \textbf{0.05} & \textbf{0.03} & 0.81 & 0.40 & 0.28 & 0.34 &\textbf{0.00} & \textbf{0.03} & 0.90 & \textbf{0.03} & 0.23 & \textbf{0.02} & 0.19 & 0.28 \\
Mental workload*                & 6 & \cellcolor[HTML]{656565}1 & \cellcolor[HTML]{c0c0c0}2 & 3 & 4 & 5 & \textbf{0.00} & 0.13 & 0.30 & 0.73 & 0.82 & 0.11 & \textbf{0.04} & \textbf{0.01} & \textbf{0.01} & 0.64 & 0.25 & 0.20 & 0.49 & 0.42 & 0.91 \\
Efficiency                      & 6 & \cellcolor[HTML]{c0c0c0}2 & 3 & 4 & 5 & \cellcolor[HTML]{656565}1 & \textbf{0.00} &\textbf{0.00} & \textbf{0.01} &\textbf{0.00} &\textbf{0.00} & 0.58 & 0.08 & 0.22 & 0.13 & 0.22 & 0.49 & \textbf{0.04} & 0.58 & \textbf{0.00} & \textbf{0.01} \\
Ambiguity*                      & 6 & \cellcolor[HTML]{656565}1 & \cellcolor[HTML]{c0c0c0}2 & 3 & 5 & 4 & \textbf{0.00} &\textbf{0.00} & \textbf{0.01} & 0.08 &\textbf{0.00} & 0.63 & 0.28 & \textbf{0.05} & 0.47 & 0.55 & 0.15 & 0.81 & 0.40 & 0.72 & 0.23 \\
Ease of use                     & 6 & \cellcolor[HTML]{656565}1 & \cellcolor[HTML]{c0c0c0}2 & 3 & 5 & 4 & \textbf{0.00} &\textbf{0.00} & \textbf{0.03} & 0.16 & \textbf{0.02} & 0.25 & \textbf{0.01} &\textbf{0.00} & \textbf{0.02} & 0.12 & \textbf{0.02} & 0.20 & 0.44 & 0.80 & 0.30 \\
Latency*                        & 6 & \cellcolor[HTML]{c0c0c0}2 & 3 & 5 & 4 & \cellcolor[HTML]{656565}1 & \textbf{0.00} & 0.10 & 0.31 & 0.13 &\textbf{0.00} & \textbf{0.01} &\textbf{0.00} & \textbf{0.01} & 0.70 & 0.53 & 0.90 & \textbf{0.03} & 0.61 & \textbf{0.01} & \textbf{0.03} \\
Visibility                      & 6 & 4 & 5 & 3 & \cellcolor[HTML]{c0c0c0}2 & \cellcolor[HTML]{656565}1 & \textbf{0.00} & \textbf{0.02} &\textbf{0.00} &\textbf{0.00} &\textbf{0.00} & 0.55 &\textbf{0.00} & \textbf{0.00} & \textbf{0.00} & \textbf{0.00} & \textbf{0.00} & \textbf{0.00} & 0.24 & \textbf{0.00} & \textbf{0.02}  \\ 

Aggregated               & 6 & \cellcolor[HTML]{c0c0c0}2 & 3 & 4 & 5 & \cellcolor[HTML]{656565}1 & \textbf{0.00} &\textbf{0.00} & \textbf{0.02} & \textbf{0.04} &\textbf{0.00} & 0.45 & 0.11 & \textbf{0.05} & 0.90 & 0.38 & 0.24 & 0.38 & 0.75 & \textbf{0.05} & \textbf{0.04} \\

\bottomrule
\end{tabular}%
}
\end{table*}

Ranking of the features analyzed in the PEQ were first analyzed. Results aggregated with the RPSS are reported (together with pairwise significances) in Table~\ref{tab:post_rank}. Considering the various features, all the interfaces were ranked significantly higher than B (apart from familiarity, as it could be expected); this finding suggests that the introduction of a vehicle-to-pedestrian interface was effective. Worth to mention are the different placements of the two AR interfaces; according to participants preferences, E significantly outperformed F for most of the features. 
The least significant differences were obtained for cognitive workload: only S was judged as significantly less demanding in terms of mental effort than B, M, F and E. In fact, S was also ranked frequently among the two best interfaces, largely overcoming E in features regarding immediateness (like ease of use and intuitiveness). 

Very similar considerations can be made based on rankings assigned in the AIQ: by performing a consistency check between the two observations (AIQ, PEQ) of the same features, all reached significant levels with high correlation ($\rho \geq 0.83$), except ambiguity ($p=0.16$) and latency ($p=0.27$). 

Like for mental demand, none of the NASA-TLX indicators (normalized to interface B on a per-sample basis) were found to be significant, with the exception of S that was considered as less demanding than B ($p=0.01$).

The best and second-best interfaces along each dimension are highlighted in Table~\ref{tab:post_rank}. It can be concluded that, while both S and E  stood out from the other interfaces, S required a lower effort for the user. Comparing the AR interfaces, participants were more effective in completing the crossing task (higher efficiency, lower latency) when using E. The blue arrow included in E could justify its  significantly higher ranking in terms of visibility. Importantly, both S and P were deemed less visible than the other interfaces (as expected), but for S this aspect was not considered detrimental to safety.

\begin{figure*}[t]
    \centering
    \includegraphics[width=1.32\columnwidth]{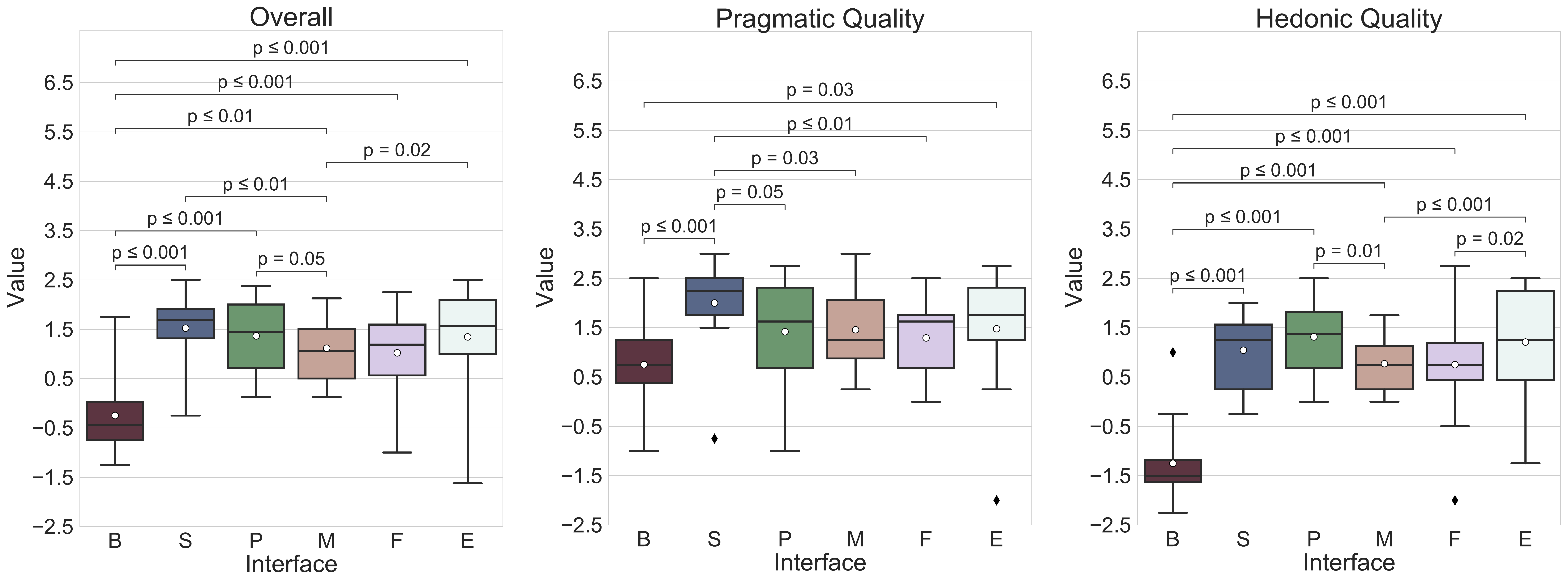}
    \caption{Results of the UEQ-S: overall, as well as pragmatic and hedonic quality dimensions shown ($p$-values reported for significant pairwise comparisons).}
    \label{fig:ueq}
\end{figure*}

Moving to the other dimensions characterising the user experience that were investigated more in depth through the UEQ-S, as reported in Fig.~\ref{fig:ueq} all the interfaces performed significantly better than B, whereas M was significantly the worst interface (although not significantly with respect to F). Analyzing separately the  dimensions addressed by the questionnaire, S was considered as significantly more pragmatic than the other interfaces, except E; moreover, focusing on the AR interfaces, only E was judged better than B in this respect. Concerning the hedonic dimension, M was judged worse than both E and P, E overcame F, and P was rated better than M.

With respect to overall usability, according to SUS results all the interfaces were rated from ``good'' to ``excellent'', and no significant differences were observed ($p=0.11$): B ($M=84.6,\,SD=10.2$), S ($M=90.6,\,SD=6.2$), P ($M=84.0,\,SD=17.8$), M ($M=79.8,\,SD=10.4$), F ($M=79.8,\,SD=13.4$), E ($M=76.1,\,SD=23.2$).

\begin{table*}[t]
\centering
\caption{Trust and safety. Means (standard deviations), and p-values for pairwise comparisons. Inverted items are marked with *.}
\label{tab:trust}
\resizebox{\textwidth}{!}{%
\begin{tabular}{lllllll|lllllllllllllll}
\toprule
                                & B & S & P & M & F & E & B-S & B-P & B-M & B-F & B-E & S-P & S-M & S-F & S-E & P-M & P-F & P-E & M-F & M-E & F-E \\ \midrule
Trust                           & 27.0(7.1) & 37.3(3.2) & 34.3(7.6) & 33.4(6.6) & 33.2(6.2) & 36.3(6.1) & \textbf{0.00} & \textbf{0.00} & \textbf{0.01} & \textbf{0.03} & \textbf{0.00} & 0.74 & 0.11 & \textbf{0.03} & 0.74 & 0.20 & \textbf{0.05} & 1.00& 0.56 & 0.20 & \textbf{0.05} \\
Safety                          & 2.3(1.0) & 4.2(1.2) & 3.9(1.1) & 3.5(0.3) & 3.4(0.6) & 4.2(0.6) & \textbf{0.00} & \textbf{0.00} & \textbf{0.00} & \textbf{0.02} & \textbf{0.00} & 0.66& 0.11 & \textbf{0.03} & 0.88 & 0.25 & \textbf{0.05} & 0.77 & 0.51 & 0.15 & \textbf{0.04} \\
Hesitancy*                      & 3.3(0.8) & 2.2(0.4) & 3.1(0.0) & 2.1(1.0) & 2.8(1.0) & 2.1(1.4) & \textbf{0.01} & 0.46 & \textbf{0.00} & 0.18 & \textbf{0.00} & \textbf{0.04} & 0.55& 0.13& 0.46 & \textbf{0.01} & 0.55 & \textbf{0.01} & \textbf{0.04} & 0.89 & \textbf{0.03}  \\
Cautiousness                         & 4.9(0.6) & 5.0(0.7) & 4.2(0.9) & 4.5(1.3) & 4.8(0.9) & 4.8(0.4) & 0.27 & \textbf{0.03} & \textbf{0.10} & 1.00 & 1.00 & \textbf{0.00} & \textbf{0.01} & 0.27 & 0.27 & 0.58 & \textbf{0.03} & \textbf{0.03} & 0.10 & 0.10 & 1.00 \\
Rashness                         & 2.0(1.0) & 3.4(1.2) & 3.3(1.2) & 3.2(1.2) & 3.0(0.4) & 3.6(1.2) & \textbf{0.00} & \textbf{0.00} & \textbf{0.00} & \textbf{0.00} & \textbf{0.00} & 0.68 & 0.54 & 0.18& 0.89 & 0.84 & 0.34 & 0.58 & 0.45 & 0.45 & 0.14 \\
\bottomrule

\end{tabular}%
}
\end{table*}

Items of the AIQ pertaining trust and safety are reported in Table~\ref{tab:trust}.
Based on the TS, all the interfaces were found to be more trustworthy than B; furthermore, S, P, and E scored significantly better than F. The same trend was observed also for the perceived safety. 
An interesting aspect to analyze is hesitancy (``I felt the need to wait for the vehicle to stop before starting to cross''): interfaces that provide an instantaneous indication of the predicted stop position (M and E) scored significantly better than those that do not offer such a feedback (B, P, F). The importance of providing the VRU with some explicit feedback that the vehicle recognized him or her (not necessarily indicating the predicted stop position) is confirmed by the fact that F did not reach significance when compared with B. Surprisingly, S obtained intermediate scores for this factor, performing significantly better than both B and P (though not  differently than M, F and E).  
Related to aspects above there is the item regarding cautiousness, intended as the participants predisposition to cross when vehicles were still far away (``I felt safe to cross when FAVs were distant''): as one could  expect, P performed significantly worse than other interfaces (S, F, E, and B), whereas S  scored unexpectedly better than M. Based on open feedback, this result could be due to the fact that the interface induced the participants  not to look at the vehicle but just at the indications shown  close to their feet. Results concerning rashness, intended as the participants predisposition to cross when vehicles were close to them, are also reported, although significance was found only for the comparisons with B. 

Remaining aspects analyzed through subjective metrics did not showed significant differences  for the various interfaces.

\subsubsection{Objective Results}

\begin{figure*}[t!]
    \centering
    \includegraphics[width=2\columnwidth]{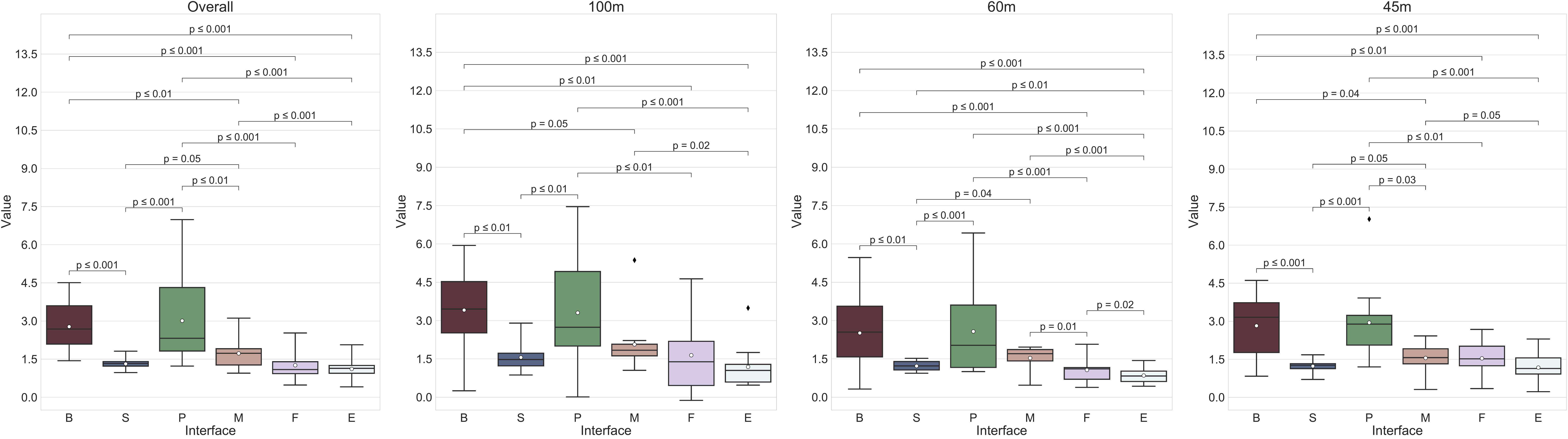}
    \caption{Decision time (lower is better): overall, and for the various distances considered ($p$-values reported for significant pairwise comparisons). }
    \label{fig:dec_t}
\end{figure*}

Data analysis on CT highlighted no significant difference among the various interfaces ($p=0.31$). Despite previous speculations, this result is not surprising, since it may suggest that, once the VRU has concluded the negotiation with the FAV (decided to cross or not), the urgency to complete the task and the trust in the FAV are predominant on his or her behavior compared to other possible information provided by the vehicle's interface.

More insights can be obtained by focusing on the distribution of DT in Fig.~\ref{fig:dec_t}. 
Considering all the distances together (overall), only P did not score significantly better than B, and both of them showed significantly higher DT values compared to the other interfaces (S, M, F, E). Furthermore, M performed significantly worse than both the AR interfaces (F, E) as well as of S, and no difference was spotted among them (S, F, E). Further considerations can be made by analyzing results for the various distances. Participants showed significantly lower DTs for M compared with B at 100 m and 45 m, but not at 60 m. Differences with respect to other vehicle-mounted interfaces tend to reduce at higher distances. This behavior can be observed for M against S (100 m) and for M against P (100 m, 60 m). The only interface able to outperform M at any distance was E. F was able to consistently perform better than M just at 60 m;  however, at that distance, F was significantly worse than E. At intermediate distances, the only interface capable of consistently offer significantly lower DTs was E. A significantly high correlation ($\rho=0.89,\,p=0.02$) was found between the overall DT and perceived latency (AIQ), confirming agreement between objective and subjective results.

\begin{figure*}[t]
    \centering
    \includegraphics[width=2\columnwidth]{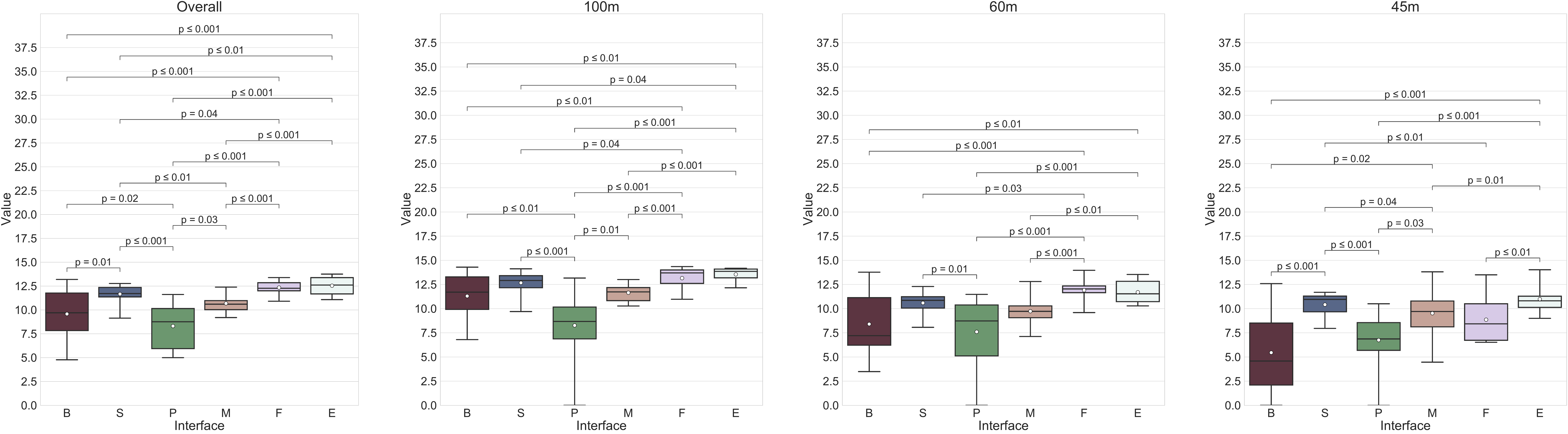}
    \caption{Speed at crossing (higher is better): overall, and for the various distances considered ($p$-values reported for significant pairwise comparisons).}
    \label{fig:speed_cx}
\end{figure*}

Another indicator that is worth discussing is the speed at which the FAV was moving when the pedestrian started crossing the road (SAC). 
Even though this speed is undoubtedly linked to DT, this is also influenced by other factors, as stated in \cite{HowShould}. It is also more unbiased that other metrics like, e.g., the DAC, which depends on the vehicle distance class. 
As shown in Fig.~\ref{fig:speed_cx}, considering all the distances (overall) the rank from best to worst is as follows: E, F, S, M, B, and P. All pairwise comparisons were significant, except E-F ($p=0.29$) and B-M ($p=0.86$). Importantly, the fact that SAC for P was even lower than for B confirms the subjective finding concerning cautiousness. Furthermore, P is the only interface for which SACs for the three distances were not statistically different ($p=0.09$). For all the other interfaces, SAC values are significantly higher at 100 m than both at 60 m and 45 m ($p \leq 0.001$). Only F had significantly degraded performance at small distances (60 m with respect to 45 m). \\
Coming back to Fig.~\ref{fig:speed_cx}, digging into the behavior of SAC for the different distances it is possible to note that the B-M, B-S, and S-M pairs gain significance merely at small distances (45 m). On the contrary, B-F, B-P, and F-P pairs gain significance at medium and large distances (60 m and 100 m). A possible interpretation for these findings could be that the closer the vehicle is, the more important is for the VRU to have a clear confirmation that it has been detected, and this relevance fades out as the distance grows. 
Moreover, although the trend for M-F could appear as contradictory, it could be easily explained by the fact that participants tended to take a confirmatory look at the vehicle for small distances after receiving the feedback from the road interface, hence delaying the time they actually started the crossing. 
Another interesting aspect is that both the AR interfaces performed better than the other interfaces (S included) at 100 m and 60 m; then, at 45 m, F fell behind both S and E, thus providing further evidence of the importance to provide VRUs with a detection feedback, especially at small distances. A possible explanation for this result could be that AR interfaces are characterized by a higher visibility, as confirmed by a correlation analysis with PEQ subjective visibility: $\rho_{100 m}=-0.89$ ($p=0.02$), $\rho_{60 m}=-0.83$ ($p=0.04$), $\rho_{45 m}=-0.77$ ($p=0.08$). It is worth observing that also SACs at 45 m were found to be significantly correlated ($\rho=0.83,\,p=0.04$) with the AIQ rashness item, suggesting that the latter could be a good metric for subjective observations. 

\begin{figure}[t]
    \centering
    \includegraphics[width=0.6\columnwidth]{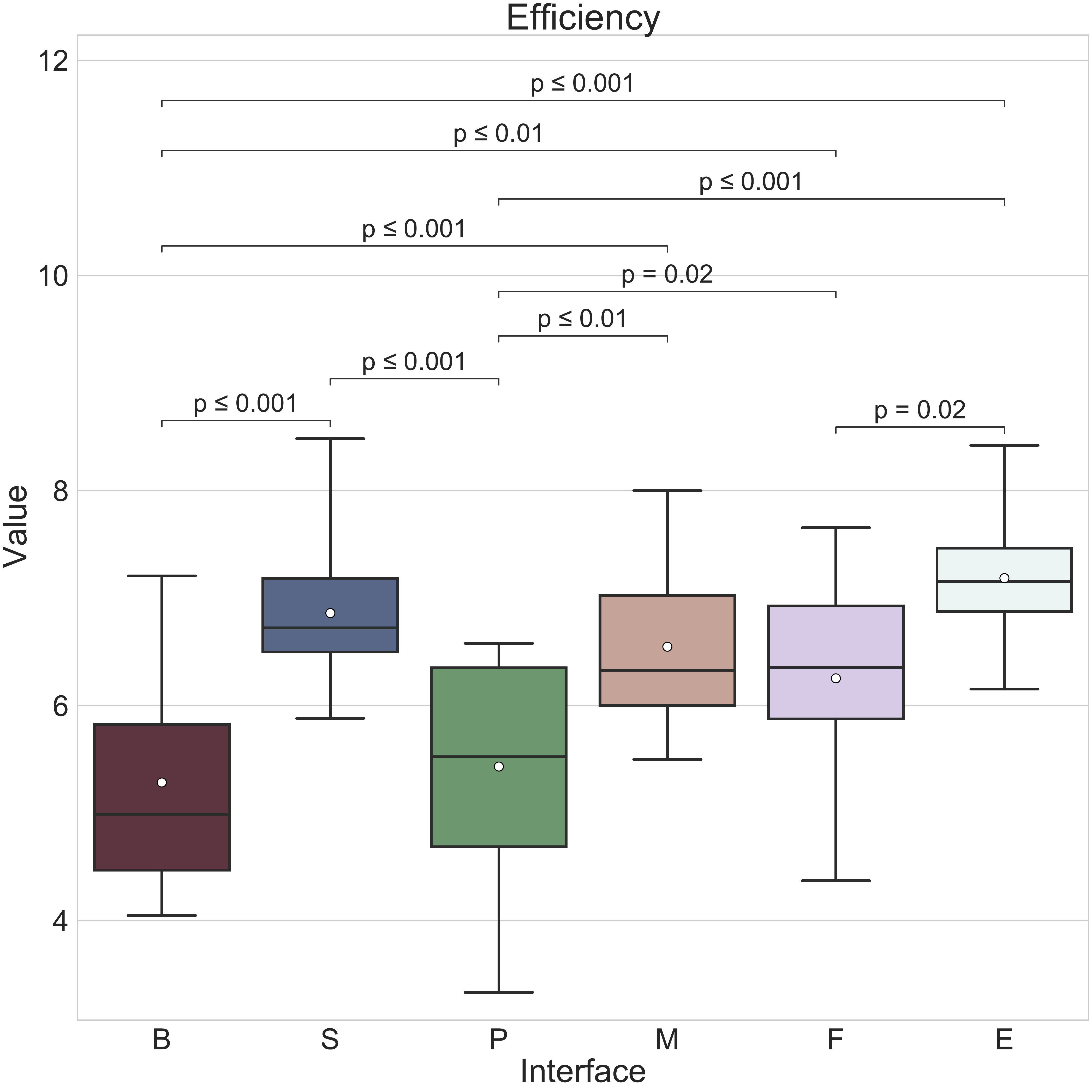}
    \caption{Objective efficiency (higher is better): no distinction was made based on distances ($p$-values reported for significant pairwise comparisons).}
    \label{fig:eff}
\end{figure}

Concerning aborted crossings and collisions, no significant differences were found among the various interfaces. 
Finally, regarding efficiency (i.e., the number of valid crossings normalized to the simulation time), Fig.~\ref{fig:eff} indicates that the most inefficient interface was P, which resulted even statistically equivalent to B. S, M, and E were found to be comparable too, whereas E appeared significantly more efficient than F. Objective efficiency was significantly correlated with the AIQ subjective efficiency ($\rho=0.83,\,p=0.04$).

\subsection{Key findings}
\label{sec:summary}
Hereafter, the key aspects of each interface as emerging from collected measurements are summarized, by considering also open feedback collected through the interviews.

\subsubsection{Smile (S)}
It proved overall to be one of two best interfaces. It is based on a very simple concept, which results in high efficiency and intuitiveness, and low mental effort. However, at large distances it is poorly visible (all the participants mentioned this issue); furthermore, $33\%$ of the participants would have also preferred it to include an additional state (``frown'') to signal when the FAV has seen them (horn activated) but cannot stop safely, and $17\%$ of them found not ideal that the predicted stop position is not indicated.

\subsubsection{Projected (P)}
It suffers as well from visibility issues; however, its main drawback is the limited efficiency, which can be attributed to its semantics. Even though it is considered very pleasant (hedonic quality), participants tend to wait for the green indication (that is showed only when the vehicle has come to a full stop); according to $50\%$ of the participants, the red crosswalk sign is not a clear indication that the FAV actually detected the pedestrian (they suggested to use the yellow color to that purpose, mimicking traffic lights). Lastly, $92\%$ of the participants found the vehicle-mounted LED panel useless or even did not notice it all.

\subsubsection{Smart roads (M)}
Even though it was judged as very familiar and visible, overall it did not score well compared to other interfaces. The main negative aspect, pinpointed by $92\%$ of the participants, was the fact that, differently than all the other interfaces, it is not visually entangled to the FAV: hence, once a change in the interface is noticed, participants tend to double-check the incoming vehicle to ensure that the interface feedback is coherent with vehicle dynamics (i.e., it is braking). Moreover, $33\%$ of the participants would have preferred an additional state indicating whether the vehicle was successfully connected to and communicating with the smart road, in order to distinguish the case of faulty pedestrian detection from faulty connections (which were not simulated).

\subsubsection{Safe roads (F)}
Although, based on objective metrics, it scored similarly to the extended version, from the subjective viewpoint it was often rated worse than the other interfaces. This result could be explained by the fact that, being the two AR interfaces very similar to each other and at the same time very different from all the others, participants may have faced psychological biases, such as the Weber--Fechner law \cite{weberfechner} and the distinction bias \cite{distinction}, making them overestimate the actual differences. The main objective difference was in the SAC at small distances, which was a clear indication of the greater safety provided by the extended version compared to the original design. By considering also poor ratings concerning mental workload, this interface is not particularly appealing compared to considered alternatives. Notwithstanding, $50\%$ of the participants expressed their appreciation for the hints provided by the green/red regions drawn on the road.

\subsubsection{Safe roads extended (E)}
It was considered as one of the two best interfaces, although more complex, mentally demanding, and less intuitive compared to the first design. Participants appreciated in particular its high visibility and the visual connection to the vehicle, as well as the indication used to simultaneously provide the pedestrian detection feedback and the predicted stop position. 
It was the only interface deemed capable of fostering trust at any distance.
It is worth remarking that $92\%$ of the participants would have preferred  fewer indications (only one participant would have liked more information), and $50\%$ of them considered not fundamental the arrows showing the vehicle's dynamics (yellow and red); $17\%$ of the participants found the red tick useful, but not the yellow arrow (as it provides a duplicated information).

\subsection{Considerations and remarks}
\label{sec:Limitations}
It is worth observing that, despite the efforts put in carrying out a fair and representative comparison, results and commentary reported above can be considered as valid only for the  scenario configured for the experiments. Different settings, encompassing, e.g., a multi-lane and/or two-way traffic could introduce important challenges for the considered interfaces, which would require further investigations. For instance, one could validate speculations that performance of interfaces providing at a glance a feedback about safe zones for the whole road width (like M, F, E) may be boosted in such a scenario. 
Similarly, one could study whether having multiple agents (either avatars controlled by the simulation or other user-controlled pedestrians) crossing the road simultaneously to the user-controlled pedestrian may lead to different communication patterns between the FAVs and the VRUs. In fact, this configuration could be extremely penalizing for interfaces conceived only for one-to-one communication like, e.g., S (as a VRU could hardy determine whether the smile was directed to him or her or to other agents). However, S could be possibly implemented in practice not as a vehicle-mounted one but as an AR interface: thus, the design would become similar to experimented solutions in which each VRU gets a different feedback for the same vehicle. In this case, future experiments should focus on digging into the possible impact of current technological limitations of wearable AR devices, e.g., related to their limited field of view. 
Still concerning the evaluation perspective, it shall be observed that the sample size of the user study reported in this work and its characteristics (e.g., the fact that participants were all Italians and accustomed to crossing in unregulated conditions, etc.) may not be fully representative of all the potential end-user categories. Hence, future works should also consider cultural factors and personal behaviors of study participants, since it can be easily expected that they could have a non-negligible impact on users' preferences.

\section{Conclusions}
\label{sec:Conclusion}

In this paper, a careful selection of the most promising state-of-the-art and newly proposed interfaces for FAV-to-VRU interaction were compared through a user study by considering a common experimental scenario represented by pedestrian crossing. Comparison was performed in an immersive VE, in which a single, user-controlled VRU was requested to cross a one-way road under non-regulated conditions. 

Results obtained using subjective and objective metrics outlined the importance to provide the users with a clear feedback about the pedestrian detection process, and to ensure high visibility. Moreover, the study proved the potential of AR-based interfaces in supporting effective vehicle-to-pedestrian communication (and, to the best of the authors' knowledge, represents the most extensive analysis of such interfaces). In particular, one of the newly proposed AR-based interfaces (namely, E) outperformed the other designs for what it concerns the above requirements; it was also characterized by a higher efficiency, and it was the only interface to be judged as capable of inducing in the users a high sense of safety independently of the vehicle distance. However, state-of-the-art designs leveraging  anthropomorphic features displayed on vehicle-mounted LED panels (like S) proved to be characterized by a lower cognitive effort and a higher intuitiveness (and ease of use, in general). Despite these findings, none of the considered interfaces stood out for all the analyzed dimensions. Nevertheless, the outcomes of this study provide precious indications that could be used to shape interaction paradigms and technologies of future FAVs ecosystems.

\section*{Acknowledgements}
The authors would like to thank Edoardo Demuru for his contribution to the system implementation. This research was partially supported by the VR@POLITO initiative. 

\ifCLASSOPTIONcaptionsoff
  \newpage
\fi




\bibliographystyle{IEEEtran}
\bibliography{cites}

\begin{thebibliography}{10}
\providecommand{\url}[1]{#1}
\csname url@samestyle\endcsname
\providecommand{\newblock}{\relax}
\providecommand{\bibinfo}[2]{#2}
\providecommand{\BIBentrySTDinterwordspacing}{\spaceskip=0pt\relax}
\providecommand{\BIBentryALTinterwordstretchfactor}{4}
\providecommand{\BIBentryALTinterwordspacing}{\spaceskip=\fontdimen2\font plus
\BIBentryALTinterwordstretchfactor\fontdimen3\font minus
  \fontdimen4\font\relax}
\providecommand{\BIBforeignlanguage}[2]{{%
\expandafter\ifx\csname l@#1\endcsname\relax
\typeout{** WARNING: IEEEtran.bst: No hyphenation pattern has been}%
\typeout{** loaded for the language `#1'. Using the pattern for}%
\typeout{** the default language instead.}%
\else
\language=\csname l@#1\endcsname
\fi
#2}}
\providecommand{\BIBdecl}{\relax}
\BIBdecl

\bibitem{mobility}
Y.~Qiao, Y.~Cheng, J.~Yang, J.~Liu, and N.~Kato, ``A mobility analytical
  framework for big mobile data in densely populated area,'' \emph{{IEEE}
  Trans. on Vehic. Techn.}, vol.~66, no.~2, pp. 1443--1455, 2016.

\bibitem{Sae}
``Taxonomy and definitions for terms related to driving automation systems for
  on-road motor vehicles,'' in \emph{SAE Technical Paper, J3016\_201806}, 2018.

\bibitem{xun2019automobile}
Y.~Xun, J.~Liu, N.~Kato, Y.~Fang, and Y.~Zhang, ``Automobile driver
  fingerprinting: A new machine learning based authentication scheme,''
  \emph{IEEE Trans. on Ind. Informatics}, vol.~16, no.~2, pp. 1417--1426, 2019.

\bibitem{networking}
J.~Wang, J.~Liu, and N.~Kato, ``Networking and communications in autonomous
  driving: A survey,'' \emph{IEEE Communications Surveys \& Tutorials},
  vol.~21, no.~2, pp. 1243--1274, 2018.

\bibitem{morra}
L.~Morra, F.~Lamberti, F.~G. Prattic{\'o}, S.~La~Rosa, and P.~Montuschi,
  ``Building trust in autonomous vehicles: Role of virtual reality driving
  simulators in {HMI} design,'' \emph{{IEEE} Trans. on Vehic. Techn.}, vol.~68,
  no.~10, 2019.

\bibitem{Rasouli_two}
A.~Rasouli and J.~K. Tsotsos, ``Autonomous vehicles that interact with
  pedestrians: {A} survey of theory and practice,'' \emph{IEEE Transactions on
  Intelligent Transportation Systems}, vol.~21, no.~3, pp. 900--918, 2019.

\bibitem{Chang}
C.-M. Chang, K.~Toda, D.~Sakamoto, and T.~Igarashi, ``Eyes on a car: {A}n
  interface design for communication between an autonomous car and a
  pedestrian,'' in \emph{Proc. 9th Int. Conf. on Automotive User Interfaces and
  Interactive Vehicular Applications}, 2017, pp. 65--73.

\bibitem{Li}
Y.~Li, M.~Dikmen, T.~G. Hussein, Y.~Wang, and C.~Burns, ``To cross or not to
  cross: Urgency-based external warning displays on autonomous vehicles to
  improve pedestrian crossing safety,'' in \emph{Proc. of 10th Int. Conf. on
  Automotive User Interfaces and Interactive Vehicular Applications}, 2018, pp.
  188--197.

\bibitem{Habibovic}
A.~Habibovic \emph{et~al.}, ``Communicating intent of automated vehicles to
  pedestrians,'' \emph{Frontiers in Psychology}, vol.~9, no. 1336, 2018.

\bibitem{Lagstrom}
T.~Lagstr{\"o}m and V.~M. Lundgren, ``{AVIP} -- autonomous vehicles'
  interaction with pedestrians -- {A}n investigation of pedestrian-driver
  communication and development of a vehicle external interface,'' Master's
  thesis, 2016.

\bibitem{HowShould}
A.~L\"{o}cken, C.~Golling, and A.~Riener, ``How should automated vehicles
  interact with pedestrians? a comparative analysis of interaction concepts in
  virtual reality,'' in \emph{Proc. 11th Int. Conf. on Automotive User
  Interfaces and Interactive Vehicular Applications}, 2019, pp. 262--274.

\bibitem{Nguyen}
T.~T. Nguyen, K.~Holl{\"a}nder, M.~Hoggenmueller, C.~Parker, and M.~Tomitsch,
  ``Designing for projection-based communication between autonomous vehicles
  and pedestrians,'' in \emph{Proc. 11th Int. Conf. on Automotive User Int. and
  Interactive Vehic. Appl.}, 2019, pp. 284--294.

\bibitem{Shinya}
S.~Kitayama, T.~Kondou, H.~Ohyabu, M.~Hirose, H.~Narihiro, and R.~Maeda,
  ``Display system for vehicle to pedestrian communication,'' in \emph{SAE
  Technical Paper}, 2017.

\bibitem{Florentine}
E.~Florentine, M.~A. Ang, S.~D. Pendleton, H.~Andersen, and M.~H. Ang~Jr,
  ``Pedestrian notification methods in autonomous vehicles for multi-class
  mobility-on-demand service,'' in \emph{Proc. 4th Int. Conf. on Human Agent
  Interaction}, 2016, pp. 387--392.

\bibitem{Lee}
Y.~M. Lee \emph{et~al.}, ``Understanding the messages conveyed by automated
  vehicles,'' in \emph{Proc. 11th Int. Conf. on Automotive User Interfaces and
  Interactive Vehicular Applications}, 2019, pp. 134--143.

\bibitem{autonomi}
L.~Graziano, ``Autono{MI} autonomous mobility interface,''
  \url{https://vimeo.com/99160686}, [Online; accessed 1/30/2021].

\bibitem{Bockle}
M.-P. B\"{o}ckle, A.~P. Brenden, M.~Klingeg{\aa}rd, A.~Habibovic, and M.~Bout,
  ``{SAV2P}: Exploring the impact of an interface for shared automated vehicles
  on pedestrians' experience,'' in \emph{Proc. of 9th Int. Conf. on Automotive
  User Interfaces and Interactive Vehicular Applications}, 2017, pp. 136--140.

\bibitem{Shuchisnigdha}
S.~Deb, L.~J. Strawderman, and D.~W. Carruth, ``Investigating pedestrian
  suggestions for external features on fully autonomous vehicles: A virtual
  reality experiment,'' \emph{Transportation Research Part F: Traffic
  Psychology and Behaviour}, vol.~59, 2018.

\bibitem{Milecia}
\BIBentryALTinterwordspacing
M.~Matthews, G.~Chowdhary, and E.~Kieson, ``Intent communication between
  autonomous vehicles and pedestrians,'' 2017. [Online]. Available:
  \url{https://arxiv.org/abs/1708.07123}
\BIBentrySTDinterwordspacing

\bibitem{Clerq}
K.~De~Clercq, A.~Dietrich, J.~P. N{\'u}{\~n}ez~Velasco, J.~de~Winter, and
  R.~Happee, ``External human-machine interfaces on automated vehicles: Effects
  on pedestrian crossing decisions,'' \emph{Human Factors}, vol.~61, no.~8,
  2019.

\bibitem{Explicit2}
\BIBentryALTinterwordspacing
E.~Ackerman, ``Drive.ai solves autonomous cars' communication problem,''
  \emph{IEEE Sprectrum}, 2016. [Online]. Available:
  \url{https://tinyurl.com/ycjw4ro2}
\BIBentrySTDinterwordspacing

\bibitem{Gruenefeld}
U.~Gruenefeld, S.~Wei{\ss}, A.~L{\"o}cken, I.~Virgilio, A.~L. Kun, and S.~Boll,
  ``{VR}oad: {G}esture-based interaction between pedestrians and automated
  vehicles in virtual reality,'' in \emph{Proc. 11th Int. Conf. on Automotive
  User Interfaces and Interactive Vehicular Applications Adjunct}, 2019.

\bibitem{Burns}
C.~G. {Burns}, L.~{Oliveira}, P.~{Thomas}, S.~{Iyer}, and S.~{Birrell},
  ``Pedestrian decision-making responses to external human-machine interface
  designs for autonomous vehicles,'' in \emph{IEEE Intelligent Vehicles Symp.},
  2019.

\bibitem{umbrellium}
\BIBentryALTinterwordspacing
J.~Mairs, ``Umbrellium develops interactive road crossing that only appears
  when needed,'' 2017. [Online]. Available: \url{https://tinyurl.com/y4vng5p8}
\BIBentrySTDinterwordspacing

\bibitem{Mahadevan}
K.~Mahadevan, S.~Somanath, and E.~Sharlin, ``Communicating awareness and intent
  in autonomous vehicle-pedestrian interaction,'' in \emph{Proc. CHI Conf. on
  Human Factors in Computing Systems}, 2018, pp. 1--12.

\bibitem{relay}
H.~Nishiyama, T.~Ngo, S.~Oiyama, and N.~Kato, ``Relay by smart device:
  Innovative communications for efficient information sharing among vehicles
  and pedestrians,'' \emph{IEEE Vehicular Technology Magazine}, vol.~10, no.~4,
  pp. 54--62, 2015.

\bibitem{mixedcann}
L.~Cancedda, A.~Cannav{\`o}, G.~Garofalo, F.~Lamberti, P.~Montuschi, and
  G.~Paravati, ``{M}ixed {R}eality-based user interaction feedback for a
  hand-controlled interface targeted to robot teleoperation,'' in
  \emph{International Conference on Augmented Reality, Virtual Reality and
  Computer Graphics}, 2017, pp. 447--463.

\bibitem{Hesenius}
M.~Hesenius, I.~B{\"o}rsting, O.~Meyer, and V.~Gruhn, ``{Don't panic!:}
  {G}uiding pedestrians in autonomous traffic with augmented reality,'' in
  \emph{Proc. 20th Int. Conf. on HCI with Mobile Devices and Services}, 2018.

\bibitem{AirSim}
S.~Shah, D.~Dey, C.~Lovett, and A.~Kapoor, ``Airsim: High-fidelity visual and
  physical simulation for autonomous vehicles,'' in \emph{Field and Service
  Robotics}, 2017, pp. 621--635.

\bibitem{frenatissime}
A.~Pillai, ``Virtual reality based study to analyse pedestrian attitude towards
  autonomous vehicles,'' 2017.

\bibitem{Gunnar}
G.~Johansson and K.~Rumar, ``Drivers' brake reaction times,'' \emph{Human
  Factors}, vol.~13, no.~1, pp. 23--27, 1971.

\bibitem{Doric}
I.~Doric \emph{et~al.}, ``A novel approach for researching crossing behavior
  and risk acceptance: The pedestrian simulator,'' in \emph{Proc. 8th Int.
  Conf. on Automotive User Int. and Interactive Vehic. Appl.}, 2016, pp.
  39--44.

\bibitem{Beggiato}
M.~Beggiato, C.~Witzlack, and J.~F. Krems, ``Gap acceptance and time-to-arrival
  estimates as basis for informal communication between pedestrians and
  vehicles,'' in \emph{Proc. 9th Int. Conf. on Automotive User Interfaces and
  Interactive Vehicular Applications}, 2017, pp. 50--57.

\bibitem{letvr}
A.~Cannav{\`o}, D.~Calandra, F.~G. Prattic{\`o}, V.~Gatteschi, and F.~Lamberti,
  ``An evaluation testbed for locomotion in virtual reality,'' \emph{{IEEE}
  Trans. Vis. Comput. Graphics}, vol.~27, pp. 1871--1889, 2021.

\bibitem{ssq}
R.~S. Kennedy, N.~E. Lane, K.~S. Berbaum, and M.~G. Lilienthal, ``Simulator
  sickness questionnaire: An enhanced method for quantifying simulator
  sickness,'' \emph{The Int. Journal of Aviat. Psychology}, vol.~3, 1993.

\bibitem{trust_scale}
J.-Y. Jian, A.~M. Bisantz, and C.~G. Drury, ``Foundations for an empirically
  determined scale of trust in automated systems,'' \emph{Int. J. Cognitive
  Ergonomics}, vol.~4, no.~1, pp. 53--71, 2000.

\bibitem{SUS}
J.~Brooke, ``{SUS} -- {A} quick and dirty usability scale,'' \emph{Usability
  Evaluation in Industry}, 1996.

\bibitem{tlx}
S.~G. Hart and L.~E. Staveland, ``Development of {NASA-TLX} ({T}ask {L}oad
  {I}ndex): {R}esults of empirical and theoretical research,'' in
  \emph{Advances in Psychology}, 1988, vol.~52, pp. 139--183.

\bibitem{ueqs}
M.~Schrepp, A.~Hinderks, and J.~Thomaschewski, ``Design and evaluation of a
  short version of the {U}ser {E}xperience {Q}uestionnaire ({UEQ-S}),''
  \emph{Int. J. Interact. Multimedia Artif. Int.}, vol.~4, no.~6, pp. 103--108,
  2017.

\bibitem{vruse}
R.~S. Kalawsky, ``{VRUSE} -- {A} computerised diagnostic tool for usability
  evaluation of virtual/synthetic environment systems,'' \emph{Applied
  Ergonomics}, vol.~30, no.~1, pp. 11--25, 1999.

\bibitem{ipq}
T.~W. Schubert, ``The sense of presence in virtual environments: A
  three-component scale measuring spatial presence, involvement, and
  realness.'' \emph{Zeitschrift f{\"u}r Medienpsychologie}, vol.~15, no.~2, pp.
  69--71, 2003.

\bibitem{rpss}
\BIBentryALTinterwordspacing
G.~Erd{\'e}lyi, L.~Piras, and J.~Rothe, ``Bucklin voting is broadly resistant
  to control,'' 2010. [Online]. Available:
  \url{https://arxiv.org/abs/1005.4115}
\BIBentrySTDinterwordspacing

\bibitem{weberfechner}
S.~Dehaene, ``The neural basis of the {W}eber--{F}echner law: {A} logarithmic
  mental number line,'' \emph{Trends in Cognitive Sciences}, vol.~7, no.~4, pp.
  145--147, 2003.

\bibitem{distinction}
C.~K. Hsee and J.~Zhang, ``Distinction bias: Misprediction and mischoice due to
  joint evaluation.'' \emph{Journal of Personality and Social Psychology},
  vol.~86, no.~5, p. 680, 2004.

\end{thebibliography}
%


\begin{IEEEbiography}[{\includegraphics[width=1in,height=1.25in,clip,keepaspectratio]{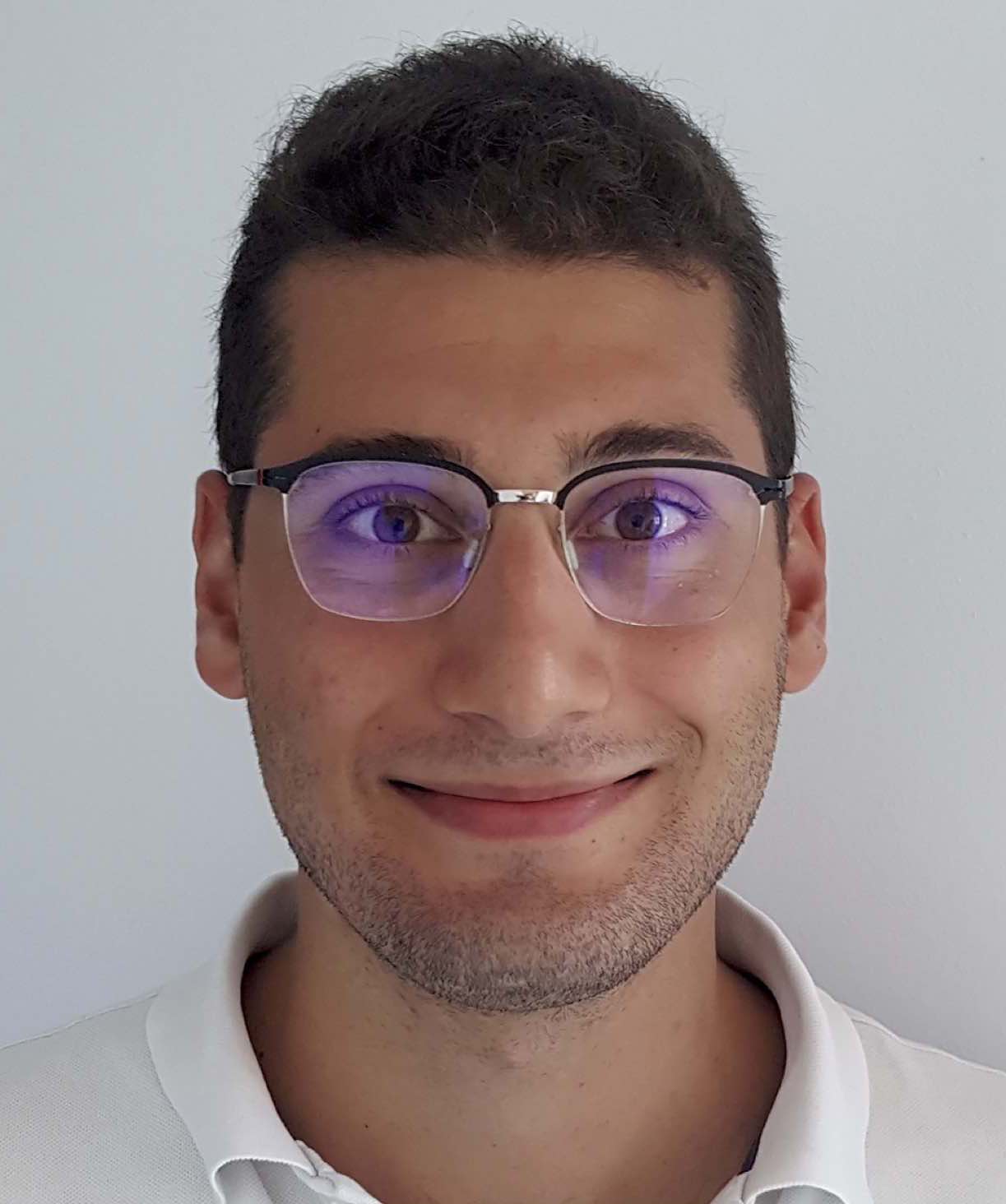}}]{F.~Gabriele~Prattic\`o}
received the M.Sc. degree in computer engineering from Politecnico di Torino, Turin, Italy, in 2017. Currently, he is a Ph.D. student at the Dipartimento di Automatica e Informatica  of  Politecnico  di  Torino, where he carries out research in the areas of extended reality, human-machine interaction, educational and training systems, and user experience design.
\end{IEEEbiography}
\vskip -2\baselineskip  plus -1fil

\begin{IEEEbiography}[{\includegraphics[width=1in,height=1.25in,clip,keepaspectratio]{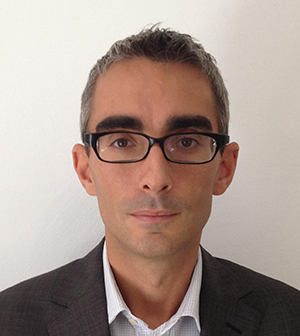}}]{Fabrizio Lamberti}
 is a Full Professor with the Dipartimento di Automatica e Informatica of Politecnico di Torino, Turin, Italy, where he has the responsibility for the VR@POLITO hub. His research interests include computer graphics, human-machine interaction, and intelligent computing. He is serving as an Associate Editor for IEEE Transactions on Computers, IEEE Transactions on Learning Technologies, IEEE Transactions on Consumer Electronics,  IEEE Consumer Electronics Magazine, and the International Journal of Human-Computer Studies. 
\end{IEEEbiography}
\vskip -2\baselineskip  plus -1fil

\begin{IEEEbiography}[{\includegraphics[width=1in,height=1.25in,clip,keepaspectratio]{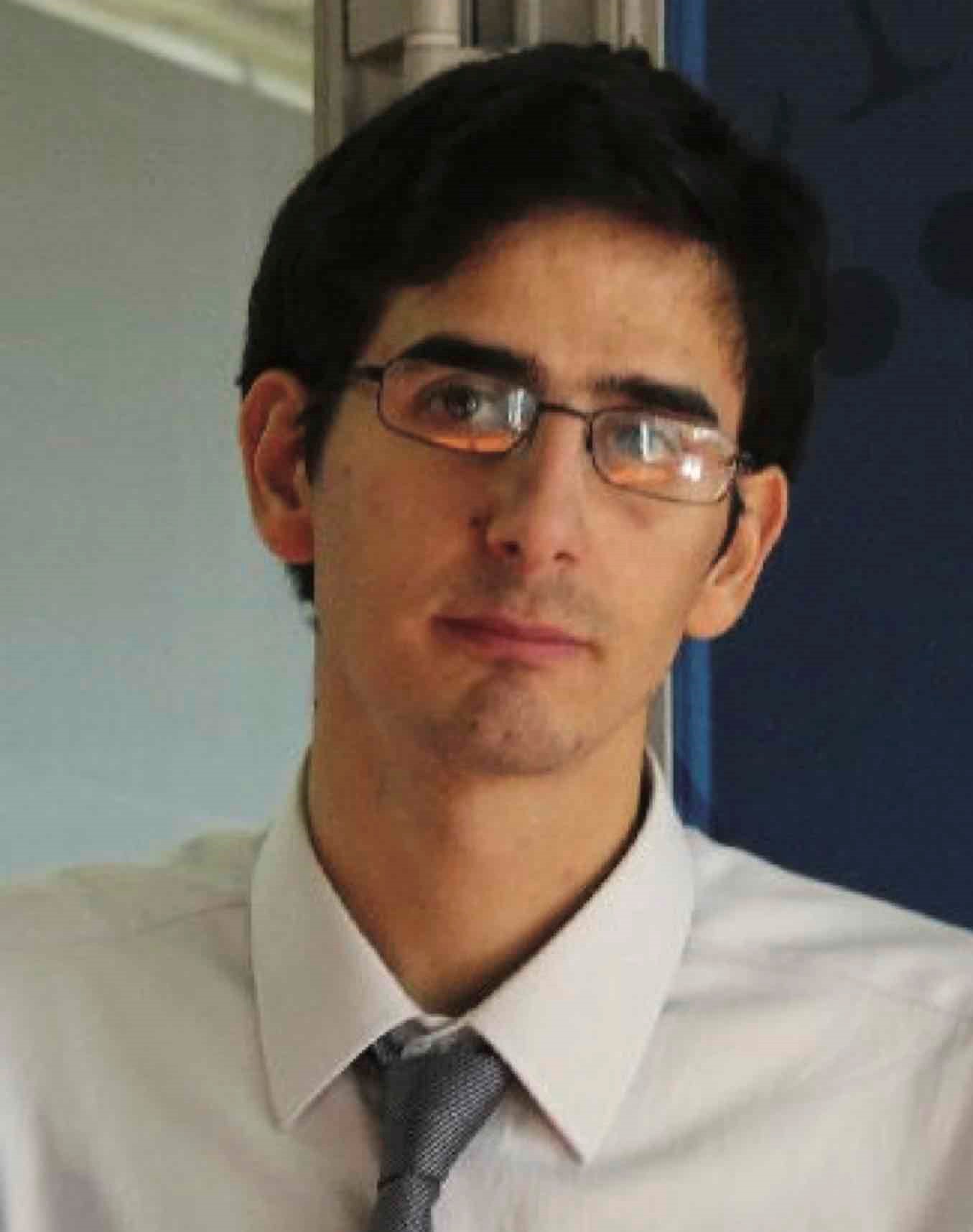}}]{Alberto Cannav\`o}
received the B.Sc. degree from University of Messina, Italy, in 2013. Then, he received the M.Sc. and the Ph.D. degrees in computer engineering from Politecnico di Torino, Italy, in 2015 and 2020, respectively. Currently, he is a Postdoctoral Fellow at the Dipartimento di Automatica e Informatica  of  Politecnico  di  Torino. His fields of interest include computer graphics and human-machine interaction.
\end{IEEEbiography}
\vskip -2\baselineskip  plus -1fil

\begin{IEEEbiography}[{\includegraphics[width=1in,height=1.25in,clip,keepaspectratio]{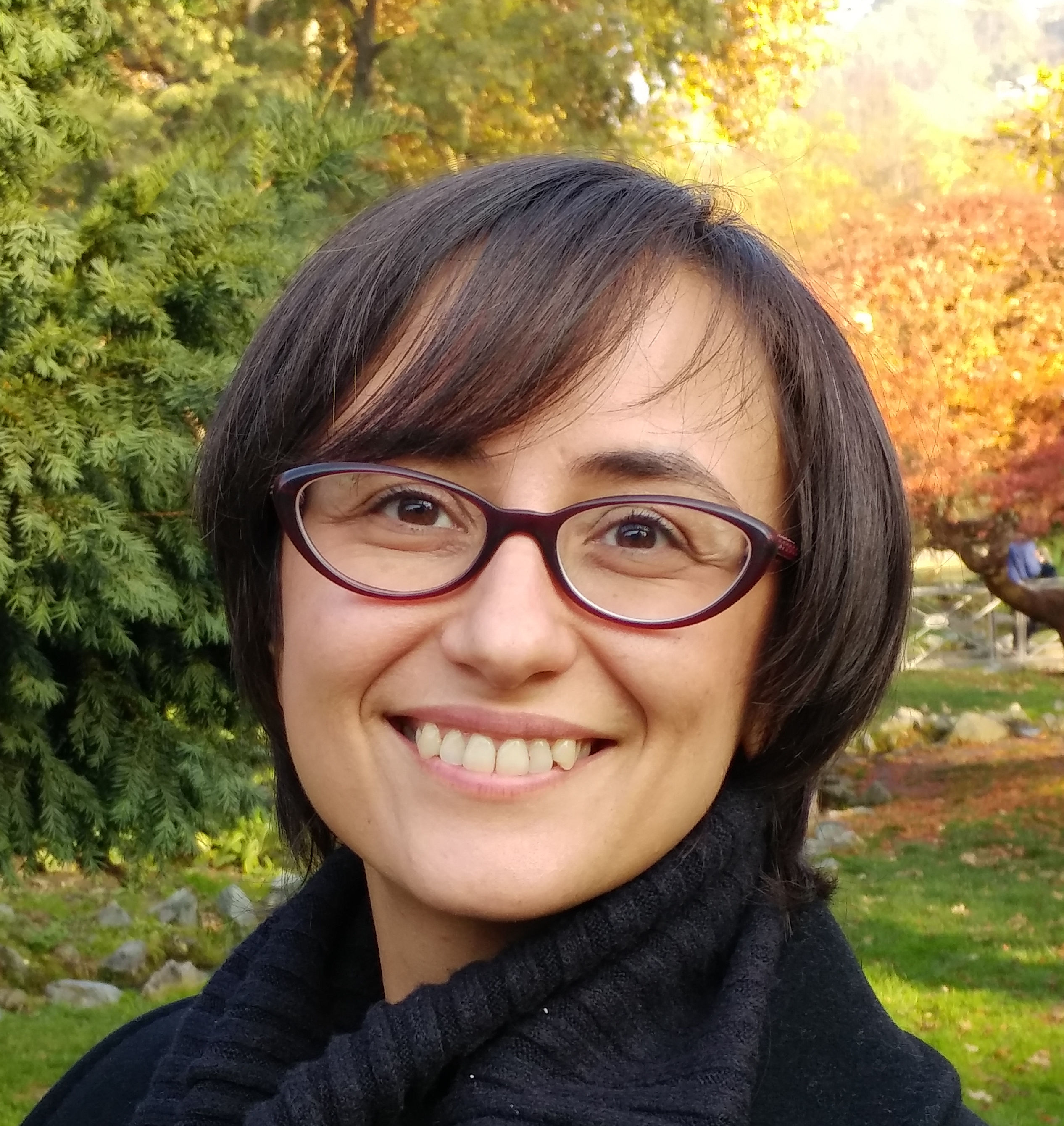}}]{Lia Morra}
received the M.Sc. and Ph.D. degrees in computer engineering from Politecnico di Torino, Turin, Italy, in 2002 and 2006, respectively. She is currently a Senior Postdoctoral Fellow with the Dipartimento di Automatica e Informatica of Politecnico di Torino. Her research interests include computer vision, pattern recognition, and machine learning.
\end{IEEEbiography}
\vskip -2\baselineskip  plus -1fil

\begin{IEEEbiography}[{\includegraphics[width=1in,height=1.25in,clip,keepaspectratio]{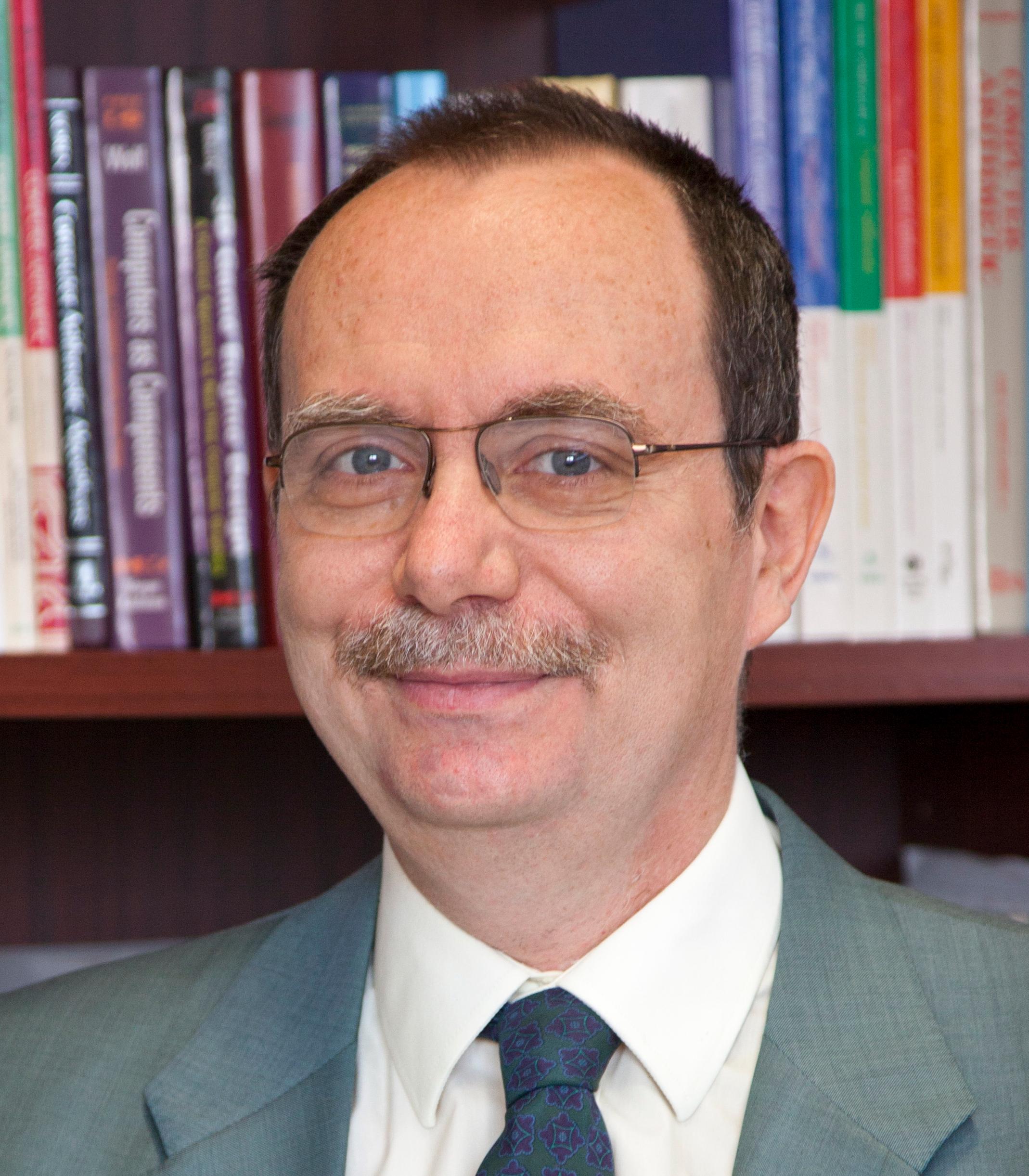}}]{Paolo Montuschi}
is a Full Professor with the Dipartimento di Automatica e Informatica and a Member of the Board of Governors of Politecnico di Torino, Italy. His research interests include computer arithmetic, computer graphics, and intelligent systems. He is serving as 2019 Acting (interim) Editor-in-Chief of the IEEE Transactions on Emerging Topics in Computing and as the 2017–2019 IEEE Computer Society Awards Chair. He is a Life Member of the International Academy of Sciences of Turin and of IEEE Eta Kappa Nu.
\end{IEEEbiography}

\vskip -2\baselineskip plus -1fil 




\enlargethispage{-5in}

\end{document}